\documentclass[preprint]{aastex61}
\accepted{April 2, 2019}
\submitjournal{ApJ}

\usepackage[]{natbib}
\usepackage{booktabs, graphicx, multirow, array, amsmath}
\usepackage{nccmath}
\usepackage{upgreek}

\def\micron{$\upmu$m}
\def\arcsec{$^{\prime\prime}$}
\def\arcmin{$^{\prime}$}
\def\nnh{N${_2}$H$^{+}$}
\def\hco{HCO$^{+}$}
\def\htco{H$^{13}$CO$^{+}$}
\def\htcn{H$^{13}$CN}

\def\amm{NH$_{3}$}
\def\amA{NH$_{3}$ (1,1)}
\def\amB{NH$_{3}$ (2,2)}
\def\amC{NH$_{3}$ (3,3)}
\def\amD{NH$_{3}$ (4,4)}

\newcommand{\Msun}{\mbox{M$_{\sun}$}}

\begin{document}

\title{Connecting the Scales: Large Area High-resolution Ammonia Mapping of NGC 1333}

\author{Arnab Dhabal}
\affiliation{Department of Astronomy, University of Maryland, College Park, MD 20742, USA}
\affiliation{NASA Goddard Space Flight Center, Greenbelt, MD 20771, USA}
\email{adhabalastro@gmail.com}

\author{Lee G. Mundy}
\affiliation{Department of Astronomy, University of Maryland, College Park, MD 20742, USA}
\affiliation{NASA Goddard Space Flight Center, Greenbelt, MD 20771, USA}
\email{lgm@astro.umd.edu}

\author{Che-yu Chen}
\affil{Department of Astronomy, University of Virginia, Charlottesville, VA 22904, USA}

\author{Peter Teuben}
\affil{Department of Astronomy, University of Maryland, College Park, MD 20742, USA}

\author{Shaye Storm}
\affiliation{Department of Astronomy, University of Maryland, College Park, MD 20742, USA}
\affiliation{Harvard-Smithsonian Center for Astrophysics, 60 Garden Street, Cambridge, MA 02138, USA}

\begin{abstract}

We use \amm\ inversion transitions to trace the dense gas in the NGC 1333 region of the Perseus molecular cloud. \amA\ and \amB\ maps covering an area of 102 square arcminutes at an angular resolution of $\sim$3.7\arcsec\ are produced by combining VLA interferometric observations with GBT single dish maps. The combined maps have a spectral resolution of 0.14~km/s and a sensitivity of 4~mJy/beam. We produce integrated intensity maps, peak intensity maps and dispersion maps of \amA\ and \amB\ and a line-of-sight velocity map of \amA. These are used to derive the optical depth for the \amA\ main component, the excitation temperature of \amA, and the rotational temperature, kinetic temperature and column density of \amm\ over the mapped area. We compare these observations with the CARMA J=1-0 observations of \nnh\ and \htco\ and conclude that they all trace the same material in these dense star forming regions. From the \amA\ velocity map, we find that a velocity gradient ridge extends in an arc across the entire southern part of NGC 1333. We propose that a large scale turbulent cell is colliding with the cloud, which could result in the formation of a layer of compressed gas. This region along the velocity gradient ridge is dotted with Class 0/I YSOs, which could have formed from local overdensities in the compressed gas leading to gravitational instabilities. The \amA\ velocity dispersion map also has relatively high values along this region, thereby substantiating the shock layer argument.

\end{abstract}


\section{Introduction}

Maps spanning large to small scales are fundamentally important to understand star formation in molecular clouds. The relative importance of processes such as turbulence, magnetic fields, gravity and chemical evolution vary at the different scales, all contributing to the birth of a protostar. In addition, feedback from star formation also affects the molecular cloud environment over a range of distances. Large area gas and dust surveys of molecular clouds ranging from parsecs to about 1000~AU are required to get the complete picture of the formation and evolution of dense structures that eventually form stars.

In going from the less dense regions of the clouds to denser filaments and on to cores, the transitions in the morphological and kinematic properties provide important clues about the underlying processes. The scaling of velocity dispersions with cloud sizes indicates that compressible turbulence plays a key role in molecular clouds \citep{Larson1981, Boldyrev2002, Henn2012}. Turbulence can affect star formation both by supporting a cloud against gravitational infall on large scales and by forming local density fluctuations that may give rise to self-gravitating cores at small scales \citep{BP1999, Padoan2002, MLKlessen2004}. Filament-like structures, that are prevalent in molecular clouds \citep{Mizuno1995, Andre2010} harbor many star forming cores. Near protostellar cores, there is increasing evidence of a sharp transition from turbulent to thermal line widths \citep{Pineda2010, Seo2015}. This behavior suggests that dense cores may form as pressure-confined structures within filaments and evolve to gravitationally bound cores before undergoing collapse to form a protostar \citep{Pattle2015, Kirk2017}. Magnetic fields within the cloud can also affect core fragmentation and collapse by providing pressure support \citep{ChenOst2012}. By comparing the kinematics of neutral and ionic species in the gas \citep{Flower2000}, it is possible to determine the presence and significance of magnetic fields. 

Radiation feedback from the embedded protostars also plays an important role in suppressing fragmentation of cores \citep{Krumholz2006}. Numerical models show that protostellar outflows and wind from intermediate-mass stars can drive turbulence in the molecular clouds \citep{Carroll2009, Offner2015}. This behavior is suggested by observations of cavities and shells in Perseus as well \citep{Quillen2005, Arce2011}. Gas temperature maps that are sensitive from large to small scales can establish the importance of the feedback processes involved in star forming regions \citep{Offner2009, Foster2009}. 

In this paper, we study the NGC 1333 region in Perseus using high angular resolution Very Large Array (VLA) observations of \amm\ inversion transitions. \amm\ is an abundant molecule that traces gas of density greater than $10^4$ cm$^{-3}$ \citep{Shirley2015} and is a late-depleter \citep{BL1997}. Nitrogen containing molecules like \amm\ and \nnh\ remain in the gas phase longer than carbon and oxygen-containing molecules such as \hco, HCN, and CO. Hence, it is particularly suitable to study cores \citep{Rosolowsky2008} and dense regions of the molecular clouds that are expected to participate in star formation in the future. Most previous observations using \amm\ involved small target areas around cores. The Green Bank Ammonia Survey (GAS) \citep{Friesen2017} is the first large-scale \amm\ survey of all the major clouds in the Gould Belt. The 32\arcsec\ beam of the GBT at 23~GHz, however, is insufficient to resolve the inner structure of cores. The VLA observations reported here make it possible to probe the regions close to the cores at a resolution of 1025~AU and study how the emission is connected to the large scale emission over the molecular cloud. 

The NGC 1333 region in Perseus is a reflection nebula at a distance of 299~$\pm$~17~pc \citep{Zucker2018}. It is known for its active low-to-intermediate mass clustered star formation \citep{Hatchell2007, Walawender2008}. There exists a wealth of information for this region, which is rich in sub-mm cores \citep{Dodds2015}, YSOs \citep{Foster2009, Johnstone2010}, outflows \citep{Plunkett2013}, Herbig-Haro objects \citep{Bally1996} and masers \citep{Lyo2014}. The \textit{Spitzer} cores to disks (c2d) \citep{Jorgensen2006, Young2015} legacy survey, the \textit{Herschel} PACS and SPIRE images \citep{Andre2010}, the \textit{James Clerk Maxwell Telescope} (JCMT) Gould Belt survey \citep{Chen2016} and the CARMA CLASSy observations \citep{Storm2014, Dhabal2018} all provide complementary data to track the YSOs and put the VLA data in larger context. 

The layout of the paper is as follows. The VLA observational set-up is discussed in Section \ref{sec:Obs}. In Section \ref{sec:Comb}, we discuss how the interferometric visibility data is combined with single-dish image data from the Green Bank Telescope (GBT) to produce maps that are sensitive to large-scale structure. Using the VLA \amA\ and \amB\ observations, in Section \ref{sec:Res} we obtain integrated intensities, peak intensities, line-of-sight velocities and velocity dispersions. In Section \ref{sec:Ana}, the primary results are used to derive optical depths, excitation temperatures, rotational temperatures, kinetic temperatures and column densities. On comparing with the CARMA \nnh\ maps and the dust maps, we establish the morphology and kinematics of the clouds in Section \ref{sec:Disc}. We also discuss some of these results in the context of filament formation and their relation to star formation. The main results are summarized in Section \ref{sec:Summ}. Appendix \ref{sec:AppA} has a brief analysis of the sensitivity of a weighting parameter involved while combining single-dish and interferometric data.

\section{VLA Observations of NGC 1333}
\label{sec:Obs}

The K-band (18-26~GHz) observations were carried out using the VLA array of 25-m antennas in D configuration providing a resolution of about 4\arcsec. The NGC 1333 region is covered by 87 pointings to cover an area of 102 square arcminutes. They were observed in multiple sessions, with 8-12 minutes integration time on each pointing. The observations were carried out from March 2013 to May 2013. 

The correlator was configured to observe the \amA, \amB, \amC\ and \amD\ lines with a channel width of 3.9~kHz (corresponding to about 0.049~km/s). The maximum sub-band bandwidth of 4~MHz was not sufficient to cover all the satellites of the ammonia inversion transitions. So two partially overlapping sub-bands were used for \amA\ and \amB. The relative intensities of the satellites are negligible for \amC\ and \amD. For these lines, we used a single sub-band covering only the main transition line. The CC$^{34}$S N=1-2, J=2-1, HNCO 1(0,1)-0(0,0), H$_2$O maser 6(1,6)-5(2,3), CCS N=1-2, J=2-1 and $^{15}$NH$_3$ (1,1) lines were also available to be observed in the same configuration (see Table \ref{tbl:Lines2}). A channel width of 7.8~kHz was used for these observations to improve the signal-to-noise ratio (SNR), anticipating their relatively low brightness temperatures. Two continuum bands of width 128~MHz were also observed at 22.3~GHz and 23.9~GHz.

Flagging and calibration were carried out using the CASA EVLA pipeline. J0336+3218, 3C84 and 3C147 were used as phase, bandpass and flux calibrators, respectively. The calibrated visibility data in the measurement sets were cleaned using the CASA \texttt{tclean} routine. The synthesized beam over the channels has a median size of 3.76\arcsec~$\times$~3.34\arcsec. For the \amm\ lines, the mean value of the spectral channel RMS noise is 6.8~mJy/beam at the native channel width. Channel binning was carried out to improve the sensitivity and look for detections in the rarer species, at the cost of velocity resolution. \amA\ and \amB\ were detected over large areas in the observed regions. \amC\ was only detected at the outflows in the NGC 1333 region. Seven 22 GHz H$_2$O masers were also visually identified from the corresponding position-position-velocity data cube. In the remaining five molecular species transitions, no emission was detected above the noise. 

\begin{table*}
\centering
\newcolumntype{P}[1]{>{\centering\arraybackslash}p{#1}}

\setlength\extrarowheight{3pt}

\begin{tabular}{ c c P{6em} P{6em} P{6em}} 
\toprule
Molecular Species & Transition & Frequency (MHz) & Channel Width (kHz) & Frequency Range (MHz)\\ 
\midrule 

NH$_3$ & (J,K)$^\pm$=(1,1)$^+$-(1,1)$^-$ & 23694.4955 & 3.9 & 5.574\\
NH$_3$ & (J,K)$^\pm$=(2,2)$^+$-(2,2)$^-$ & 23722.6333 & 3.9 & 5.574\\
NH$_3$ & (J,K)$^\pm$=(3,3)$^+$-(3,3)$^-$ & 23870.1292 & 3.9 & 1.997\\
NH$_3$ & (J,K)$^\pm$=(4,4)$^+$-(4,4)$^-$ & 24139.4163 & 3.9 & 1.997\\
$^{15}$NH$_3$ & (J,K)$^\pm$=(1,1)$^+$-(1,1)$^-$ & 22624.9295 & 7.8 & 1.997\\
HNCO & J(K$_a$,K$_c$)=1(0,1)-0(0,0) & 21981.5726 & 7.8 & 1.997\\
H$_2$O & J(K$_a$,K$_c$)=6(1,6)-5(2,3) & 22235.0798 & 7.8 & 1.997\\
CCS & N=1-2, J=2-1 & 22344.0308 & 7.8 & 1.997\\
CC$^{34}$S & N=1-2, J=2-1 & 21930.4860 & 7.8 & 1.997\\
\bottomrule
\end{tabular}

\caption{Observed Molecular Lines. The transtion frequencies are obtained from the CDMS \citep{CDMS2005}.}
\label{tbl:Lines2}
\end{table*}

\section{Combination of Interferometric and Single Dish Data}
\label{sec:Comb}

The maps produced from the VLA data alone have strong negative sidelobes, which is typical of maps produced from interferometric visibility data due to the lack of small u-v spacings \citep{Kogan2000}. In our VLA data, the minimum u-v spacing is 25~m. Consequently, in many areas of the map the negative sidelobes corresponding to strong emission elsewhere suppress the actual flux levels and spatial variations even though the interferometer is sensitive to small scale structure. This issue can be remedied by using complementary single dish observations \citep{Vogel1984}. The NGC 1333 region has been observed in \amA, \amB\ and \amC\ with the Green Bank Telescope (GBT) as part of the Green Bank Ammonia survey (GAS) \citep{Friesen2017} and the data are publicly available \citep{GBT2017}. These data were combined with the high resolution VLA maps to make the maps sensitive to large scale structure as well. In the rest of the paper, we only discuss the \amm\ inversion transition observations in the NGC 1333 region.

The strategy described in \cite{Koda2011} is adopted to combine the interferometric and single dish data. This process involves converting the single dish data to visibilities to fill the inner part of the u-v plane, followed by weighting these visibilities with respect to the interferometric visibilities. This combined visibility dataset is then inverse Fourier transformed and cleaned to produce the final image. There are multiple ways to combine the data from single dish telescopes and interferometers. Other methods involve combining the data in the image domain \citep{Helfer2003}, or further Fourier transforming the image domain data and then combining them \citep{Weiss2001} (as is done in the CASA \texttt{feather} routine). Since inverse Fourier transforming the visibilities to produce the interferometric image is a non-linear process, combining them after this step leads to loss of information from the interferometric component. Thus we combine the data in the visibility domain.  

The GBT single dish maps have a spectral resolution of 5.7~kHz ($\sim$~0.07~km/s at 23.7~GHz). We binned the channels by a factor of 2 to improve the SNR, while retaining 5 channels to cover the typical FWHM linewidth of 0.7~km/s. The resulting channel width is about 3 times the VLA channel width. A total of 400 channels cover all the \amA\ and \amB\ hyperfine components. We converted the map from K to Jy/beam using a factor of 0.69 taking into account the GBT beam size of 31.8\arcsec~$\times$~31.8\arcsec. The mean RMS of the resulting SD map is 0.059~Jy/beam. 

After extracting the relevant region of the map, we generate fake visibility data from the single dish data cubes using the \texttt{tp2vis} program \citep{Koda2011}. This involves sampling the GBT single dish map at the VLA pointings with a 25~m primary beam (same as the VLA beam) and with u-v distances up to 100~m (equivalent to the GBT diameter). The single dish visibilities are inverse Fourier transformed to check how the resulting map compares with the input single dish map. The flux levels are within 10\% of original map, and there are no noticeable differences in the intensity distributions. 

\begin{figure*}
\centering
\includegraphics[width=\textwidth]{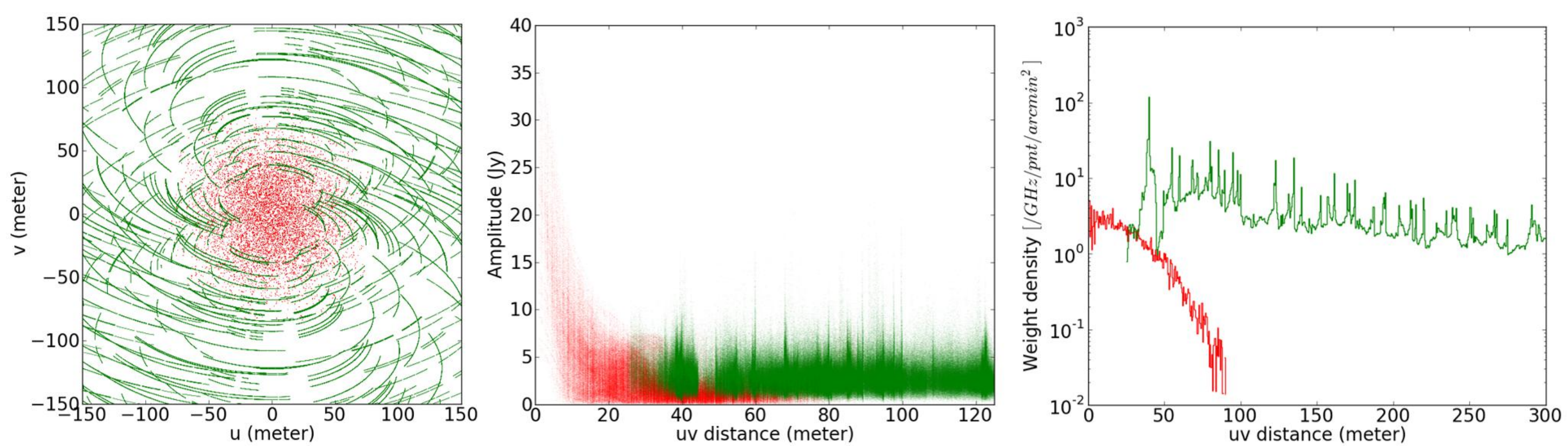}
\caption{\textit{Left}: u-v plane showing the coverage of VLA observations (green) up to 150~m, and the coverage of the visibilities generated from the GBT observations (red). \textit{Middle}: Comparison of the GBT (red) and VLA (green) visibility amplitudes. \textit{Right}: Matched weights of GBT (red) and VLA (green) visibilities after scaling.}
\label{fig:SDIntVis}
\end{figure*}

The weight densities (weights/GHz/pointing/arcmin$^2$) of the single dish visibilities are compared with those of the interferometric visibilities. Accordingly, the single dish weights are scaled to roughly match them (Figure \ref{fig:SDIntVis}). This ensures that the sensitivities transition smoothly from the larger scales to the smaller scales. Further analysis of the effect of weight scaling on the resulting maps is carried out in Appendix \ref{sec:AppA}. The combined set of visibilities is now used to generate the dirty map and cleaned using the \texttt{qac\_clean} routine that uses the CASA \texttt{tclean} routine \citep{Teuben2018}. We use the multiscale deconvolver with scales of 0\arcsec, 4\arcsec\ and 12\arcsec. The cleaning proceeds by a set of major cycles transforming between visibility and image domains while progressively updating the model and residual images. Within each major cycle, an inner loop of minor cycles cleans the map in the image domain. We needed to induce more major cycles without cleaning too deep in the corresponding minor cycles to avoid picking up interferometric artifacts. So a `cycleniter' parameter of 1000 was used. Based on the noise RMS of 0.004~Jy/beam, we also use a `threshold' parameter value of 0.02~Jy/beam, which decides the minimum target peak flux at the end of each minor cycle for each channel. Typically, the channels containing signal are cleaned in 10-15 major cycles. In regions having strong emission, the flux remaining in the residuals is less than 10\% of the total flux. 

\begin{table*}
\centering
\setlength\extrarowheight{3pt}


\begin{tabular}{ c c c c c} 
\toprule
Images & u-v range & Channel Width$^a$ & Synthesized Beam & Sensitivity\\ 
\midrule 

VLA & 25 - 1000 m & 3.9 kHz & 3.76\arcsec $\times$ 3.34\arcsec & 0.007 Jy/beam\\
GBT & 0 - 100 m & 5.7 kHz & 31.8\arcsec $\times$ 31.8\arcsec & 0.083 Jy/beam\\
Combined & 0 - 1000 m & 11.4 kHz & 3.93\arcsec $\times$ 3.41\arcsec & 0.004 Jy/beam\\
\bottomrule
\end{tabular}

\raggedright{\footnotesize{$^a$ Native channel widths for VLA and GBT maps, and rebinned channel width for combined map.}}
\caption{Properties of the original and combined NGC 1333 images}
\label{tbl:GBTVLA}
\end{table*}

\begin{figure*}
\centering
\includegraphics[trim={2.5cm 2.55cm 2.3cm 3.6cm},clip,width=\textwidth]{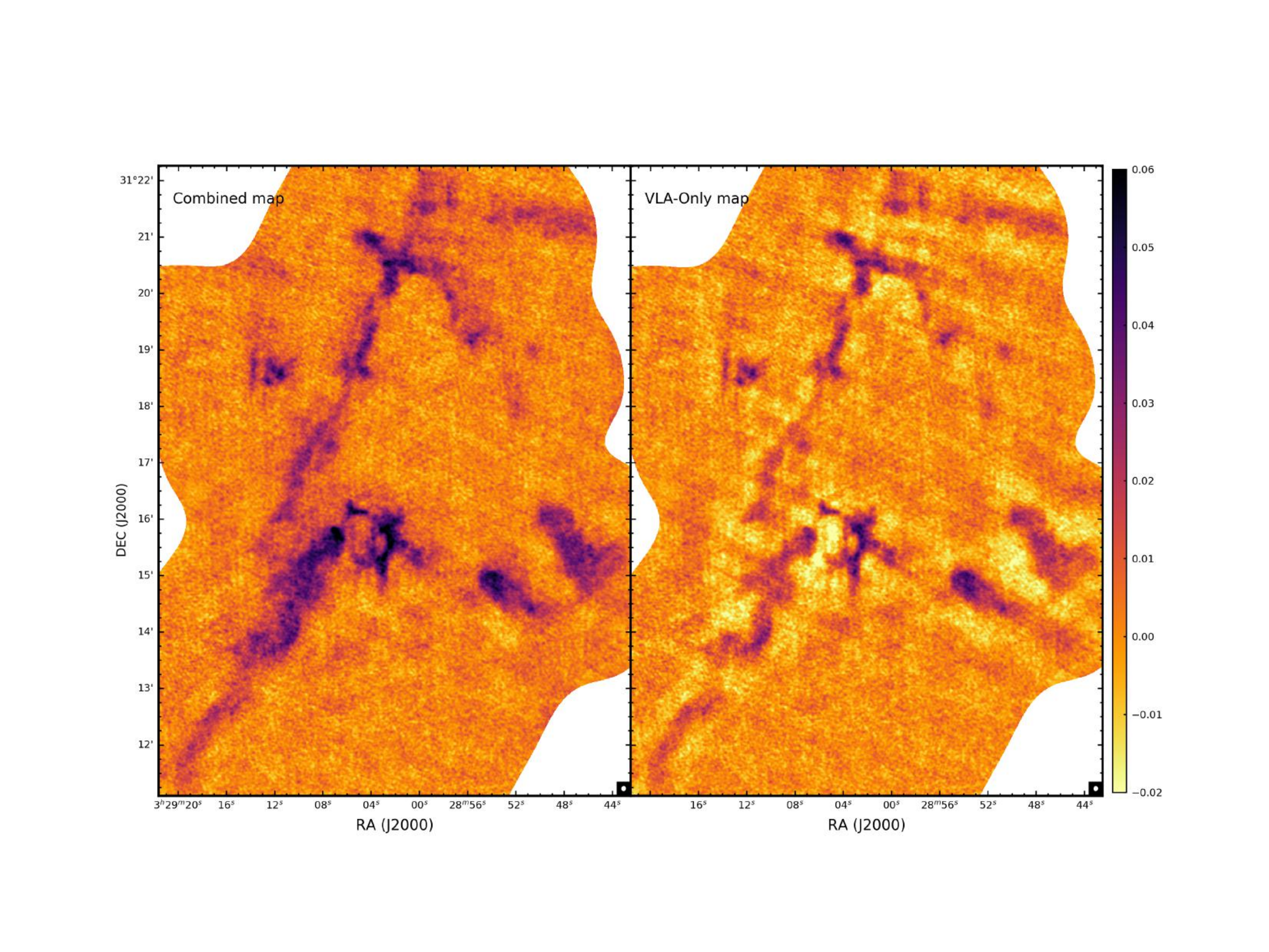}
\caption{Comparisons of the combined VLA and GBT map (left) with VLA-only map (right) shown here for a single channel (8.11-8.25~km/s) after cleaning. Small scale flux variations are present in both, but the distinct negative sidelobes of the VLA only map are absent in the combined map. The colorbar is in Jy/beam.}
\label{fig:IntComp}
\end{figure*}

\begin{figure*}
\centering
\includegraphics[trim={2.5cm 2.55cm 2.3cm 3.6cm},clip,width=\textwidth]{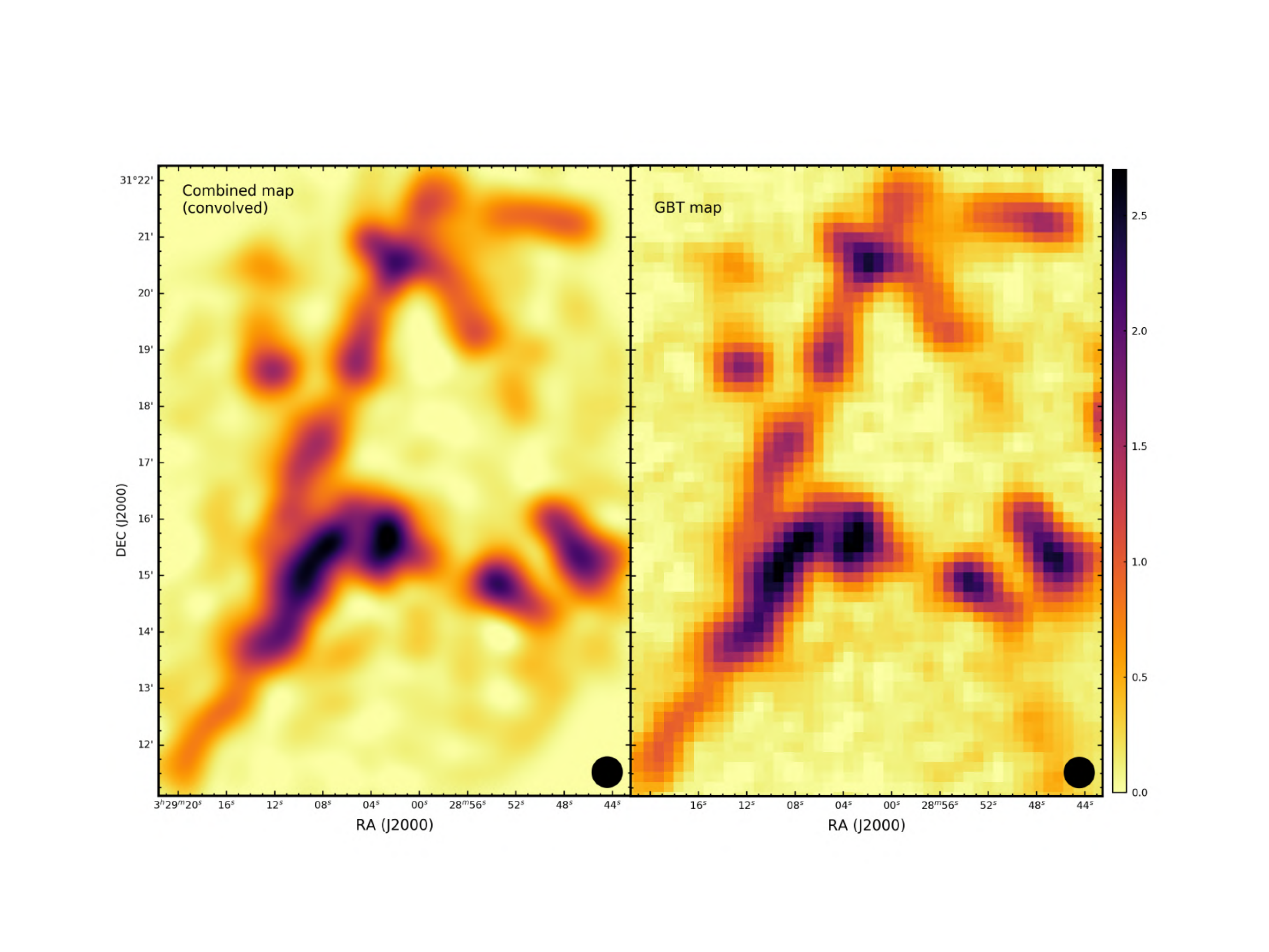}
\caption{Comparisons of combined VLA and GBT map convolved to the GBT beam size (left) with the original GBT map (right) for a single channel (8.11-8.25~km/s). The fluxes match up with less than 5\% variations in most regions. The colorbar is in Jy/beam.}
\label{fig:SDComp}
\end{figure*}

The final maps thus obtained do not have negative sidelobes and show the high resolution features, while being sensitive to large scale structure. The synthesized beam is 3.93\arcsec~$\times$~3.41\arcsec. The flux values are compared to the interferometric maps obtained from the VLA data alone in Figure \ref{fig:IntComp}. Typically, the average flux in regions of strong emission in the VLA maps is about 20\% lesser than the corresponding flux in the combined maps, but the local spatial variations of the flux are very similar. The maps are also smoothed to the GBT beam size and compared to the original GBT map (Figure \ref{fig:SDComp}). On binning the combined map to the GBT-only map pixel size, the total flux matches to about 1.5\%, while the variability in scales of the GBT beam is about 5\%. Since the \amB\ emission traces warmer gas, it is less extended, and consequently the flux levels in the combined map are better matched with the interferometric map fluxes than for \amA.

\section{Results}
\label{sec:Res}

The \amm\ inversion transitions involve quantum tunneling of the N nuclei through the plane of the H nuclei in the same rotational quantum state (J,K). The inversion transitions however have hyperfine components due to electric quadrupole coupling (quantum number denoted by F$_1$) and magnetic dipole coupling (quantum number denoted by F). In the case of \amA, there are 18 hyperfine components that are bunched into 5 groups of 2, 3, 8, 3 and 2 lines based on the F$_1$ transition. Their theoretical relative intensities are 0.111, 0.139, 0.5, 0.139 and 0.111, respectively. Similarly, the \amB\ transition has 24 hyperfine components also divided into 5 groups of 3, 3, 12, 3 and 3 lines, having relative intensities 0.05, 0.052, 0.796, 0.052 and 0.05, respectively \citep{Ho1983, Magnum2015}. In both cases, the central line is the main component, and the other four components are called satellite components. 

\subsection{Ammonia (1,1) Results}
\label{amm11}

\begin{figure*}
\centering
\includegraphics[trim={1.45cm 1.25cm 2.45cm 2.5cm},clip,width=\textwidth]{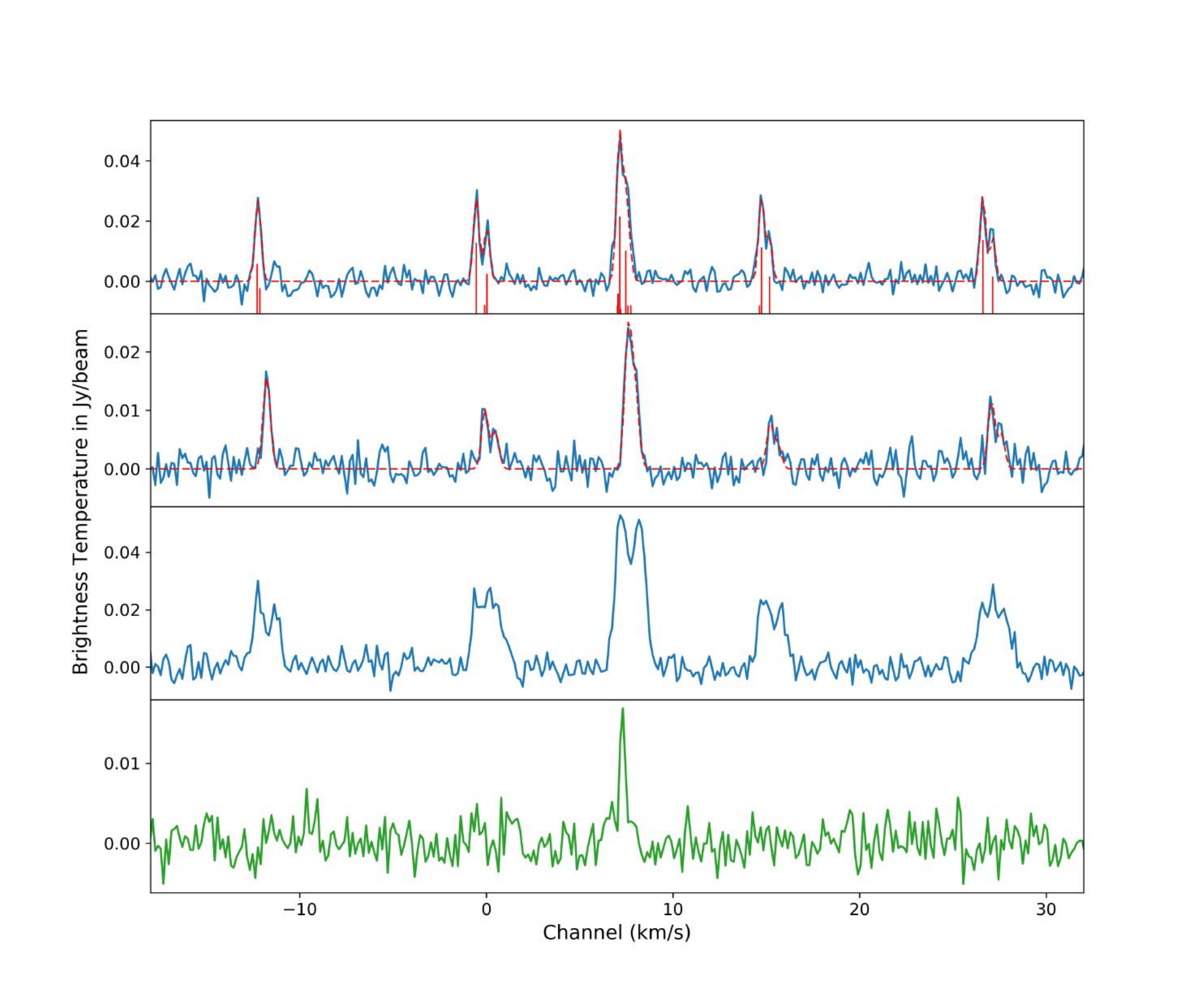}
\caption{\textit{Top row:} \amA\ spectrum at the West region, where the relative intensities of the satellite components are similar. The dotted line gives the spectral fit, while the vertical lines on the x-axis indicate the relative frequencies and the amplitudes of the 18 hyperfine components. Three of the satellite components are resolved into two peaks because of the hyperfine components. \textit{Second row:} \amA\ spectrum at an area in the SVS 13 region, where the intensities of the satellite components deviate from the theoretical expectations. The inner satellites have much lesser intensities compared to the outer ones. The dotted line indicates the spectral fit. \textit{Third row:} \amA\ spectrum at an area also in the SVS 13 region, showing the presence of multiple velocity components. The difference in the line-of-sight velocities is greater than 1~km/s. \textit{Bottom row:} \amB\ spectrum at the same area as the \amA\ spectrum in the top-most row.}
\label{fig:SpectAmm}
\end{figure*}

\begin{figure*}
\centering
\includegraphics[trim={2.55cm 2.0cm 2.9cm 3.0cm},clip,width=0.82\textwidth]{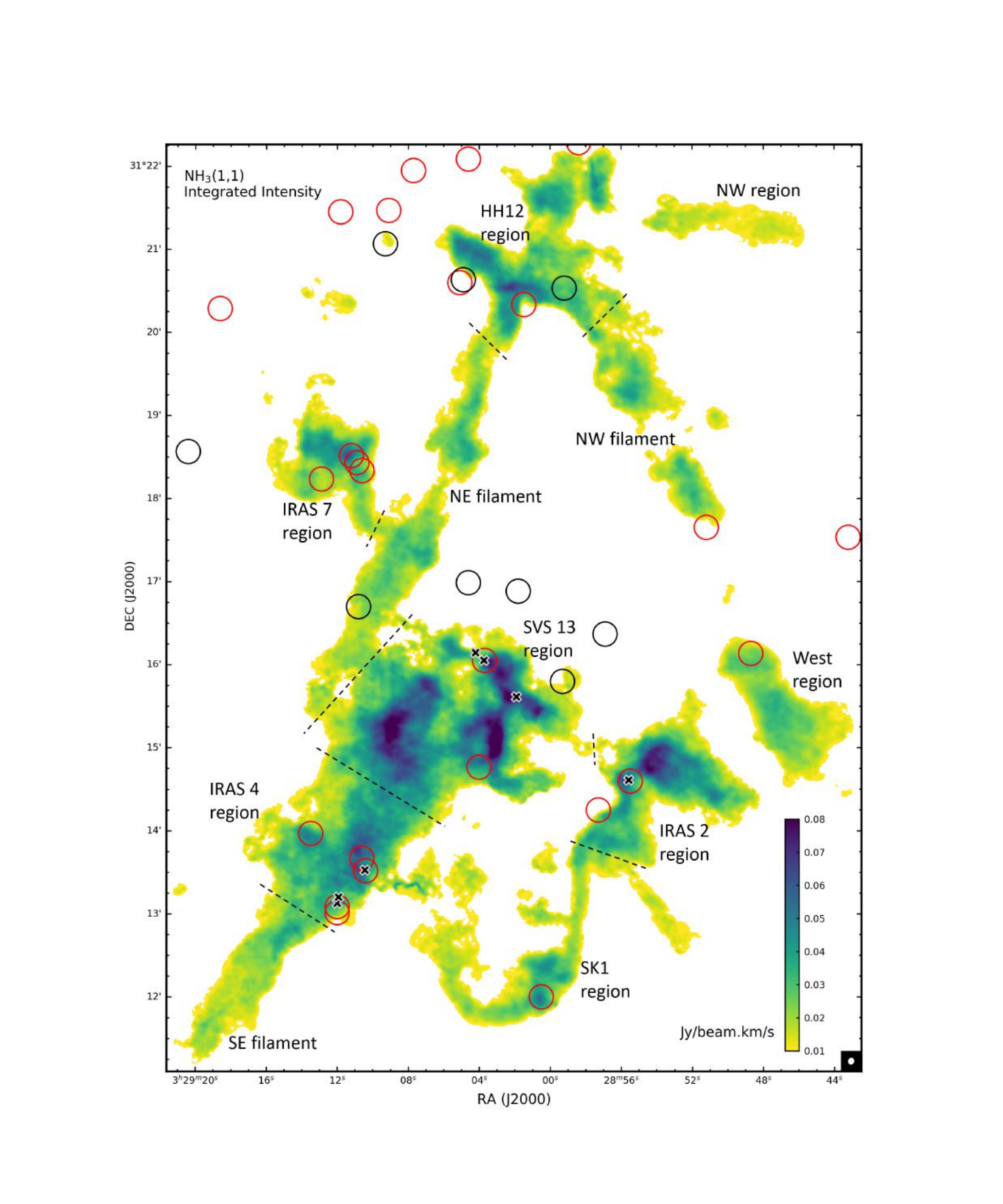}
\caption{\amA\ (main component) integrated intensity map for the NGC 1333 region. The map beam size is indicated at the bottom-right corner. The typical uncertainty is 0.005~Jy/beam~km/s. The locations of the known Class 0/I (red circles) and Flat (black circles) YSOs from \textit{Spitzer} observations \citep{Cat2015} are marked. The circles correspond to a diameter of roughly 5000~AU (about 5 times the map resolution). The crosses (`x') mark the seven water masers identified from our data. The different regions discussed in the paper are also indicated. They are separated by dotted lines in regions of continuous emission.}
\label{fig:AmmM0}
\end{figure*}

\amA\ main and satellite lines are detected over large areas in the NGC 1333 region. The relative intensities of the hyperfine components, however, vary in different regions from their theoretical relative intensities (see top two panels of Figure \ref{fig:SpectAmm}). In many regions, we also note the presence of two velocity components in the same line-of-sight. The velocity peaks are separated by as much as 1.5~km/s in some locations, indicating that the two velocities belong to independent components (for example, see third panel of Figure \ref{fig:SpectAmm}). Integrated intensity maps are generated from the position-position-velocity data cubes, by using channels near the line centers containing signal, and clipping them at 3\,$\sigma$ based on the RMS noise of each channel map. Figure \ref{fig:AmmM0} shows the integrated intensity map for the main \amA\ component alone. We have roughly divided the map into 11 regions based on their location in the map or the main embedded protostellar source(s) in the region. 

IRAS 4, IRAS 2 and SVS 13 are well-known YSO systems, each of which has been resolved into multiple sources \citep{Walawender2008}, and are all located in the central cloud core. There is a known Herbig-Haro object HH 12 in the northern region, and it is located at the apex of the two filaments (NW and NE filaments). To the west of the HH 12 region, there is a region of modest emission (NW region), but no protostars are detected there. There is a similar relatively quiescent area (West region), about 1\arcmin\ west of the IRAS 2 region. The IRAS 7 protostellar cluster region is located to the east of the NE filament. Towards the south, there is a narrow `u' shaped filament connecting the IRAS 2 and the IRAS 4 regions. At the bottom of the `u' is a known Class 0 YSO source -- SK 1 \citep{Dionatus2017}. There is another narrow short filament in this area extending towards the south-west from the IRAS 2 region. The cloud core near IRAS 4 extends to a wide filament in the south-east, which has been studied in \cite{Dhabal2018} as part of the CLASSy-II regions.  

\begin{figure*}
\centering
\includegraphics[trim={2.5cm 2.0cm 2.9cm 3.0cm},clip,width=0.85\textwidth]{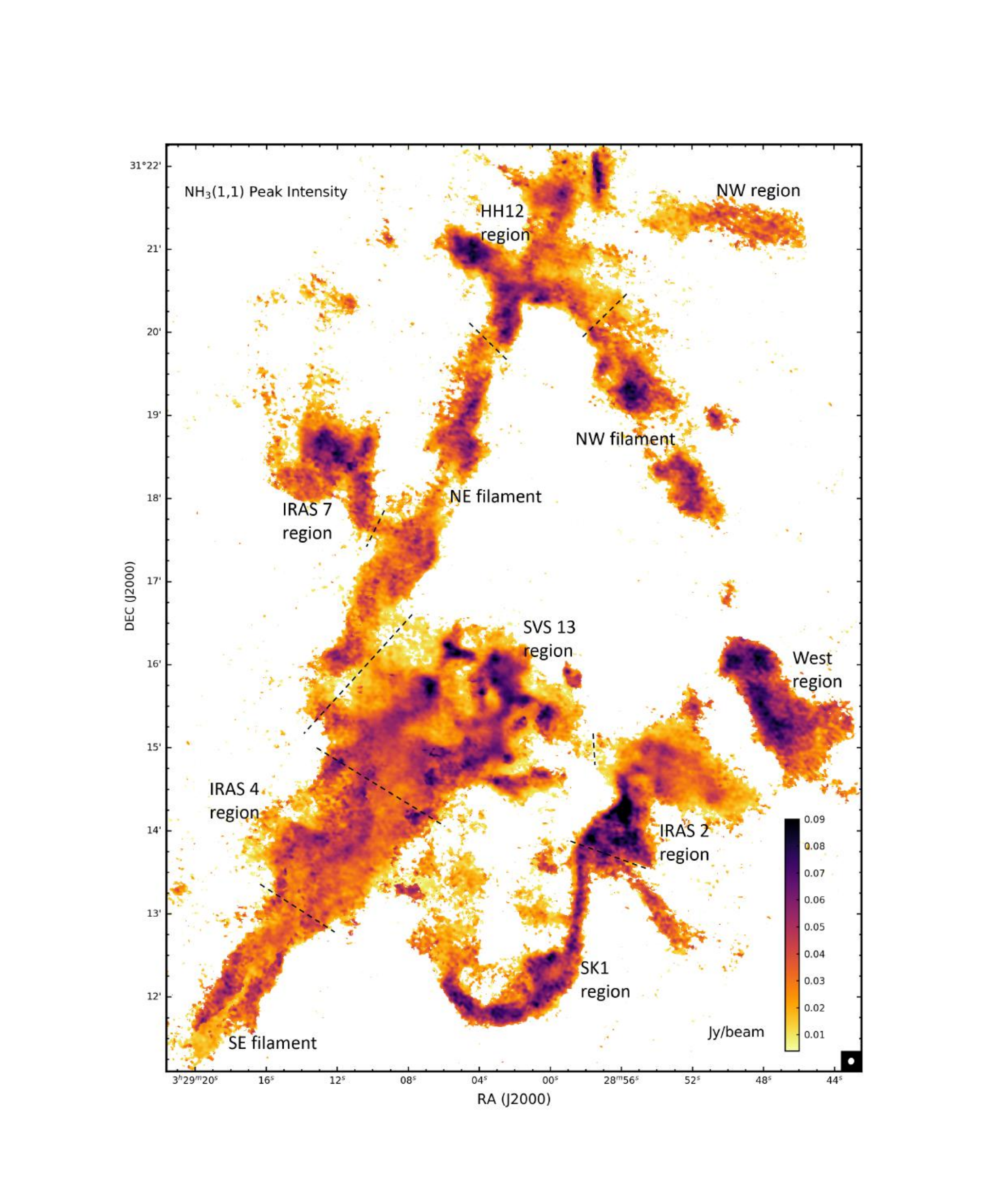}
\caption{\amA\ main component peak intensity map for the NGC 1333 region. The typical uncertainty is 0.0029~Jy/beam.}
\label{fig:MainAmp}
\end{figure*}

\begin{figure*}
\centering
\includegraphics[trim={0.5cm 1.0cm 1.4cm 1.6cm},clip,width=0.85\textwidth]{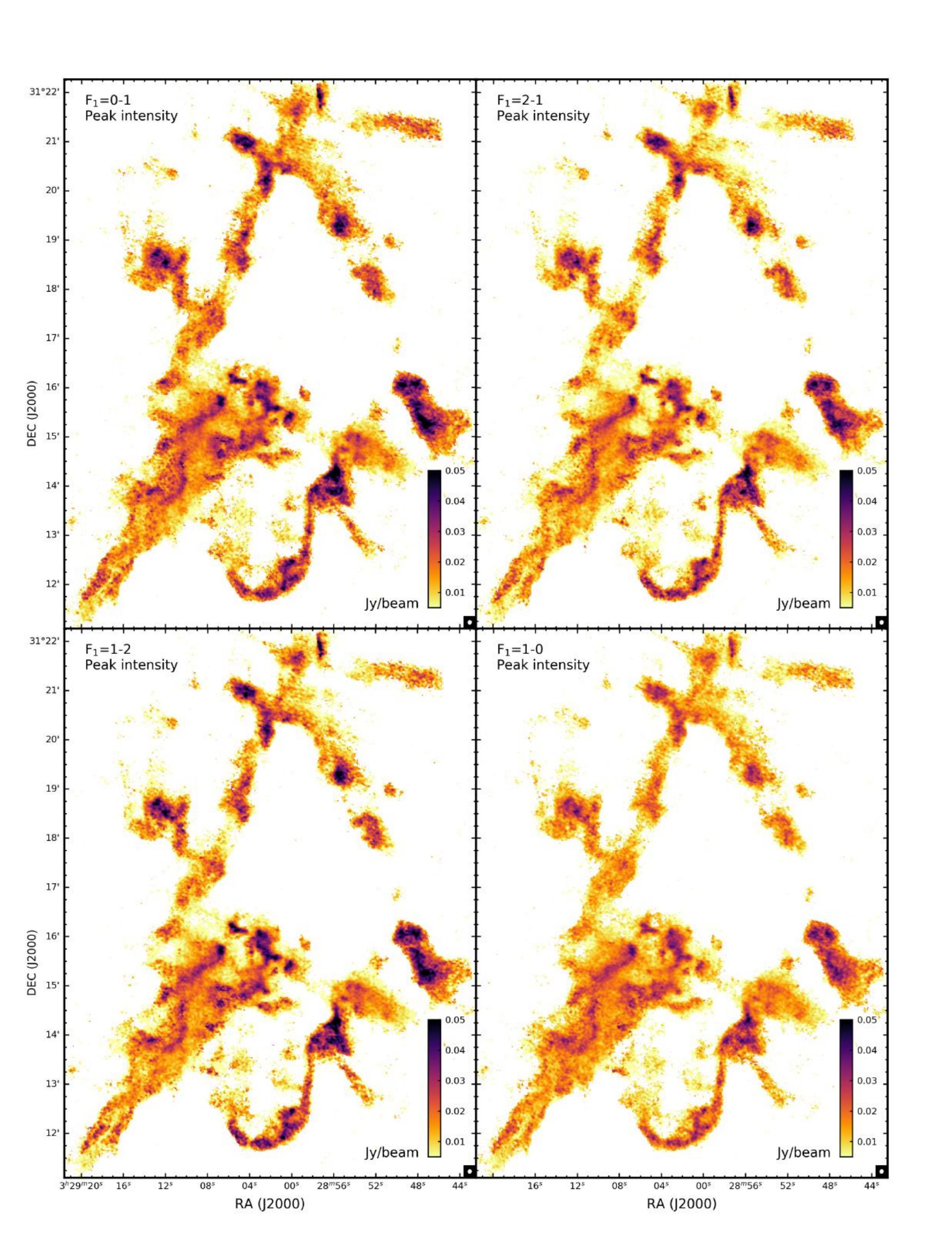}
\caption{\amA\ satellite components peak intensity map for the NGC 1333 region obtained by spectral fitting. The typical uncertainties are about 0.0024~Jy/beam.}
\label{fig:SatAmp}
\end{figure*}

\begin{figure*}
\centering
\includegraphics[trim={2.5cm 2.0cm 2.9cm 3.0cm},clip,width=0.85\textwidth]{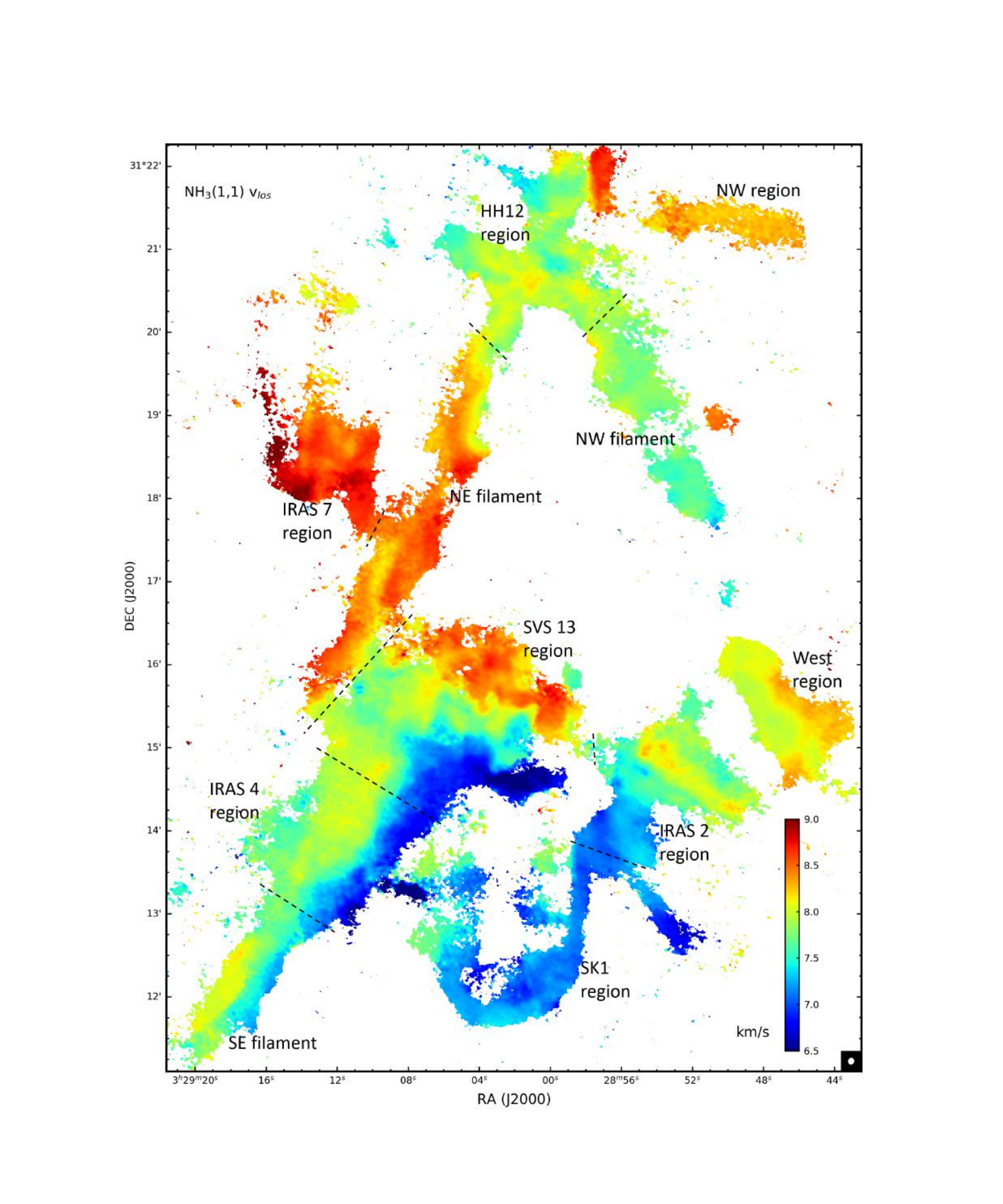}
\caption{NGC 1333 line-of-sight velocity map from spectral fitting of \amA. The typical uncertainty is 0.04~km/s.}
\label{fig:AmmVel}
\end{figure*}

\begin{figure*}
\centering
\includegraphics[trim={2.5cm 2.0cm 2.9cm 3.0cm},clip,width=0.85\textwidth]{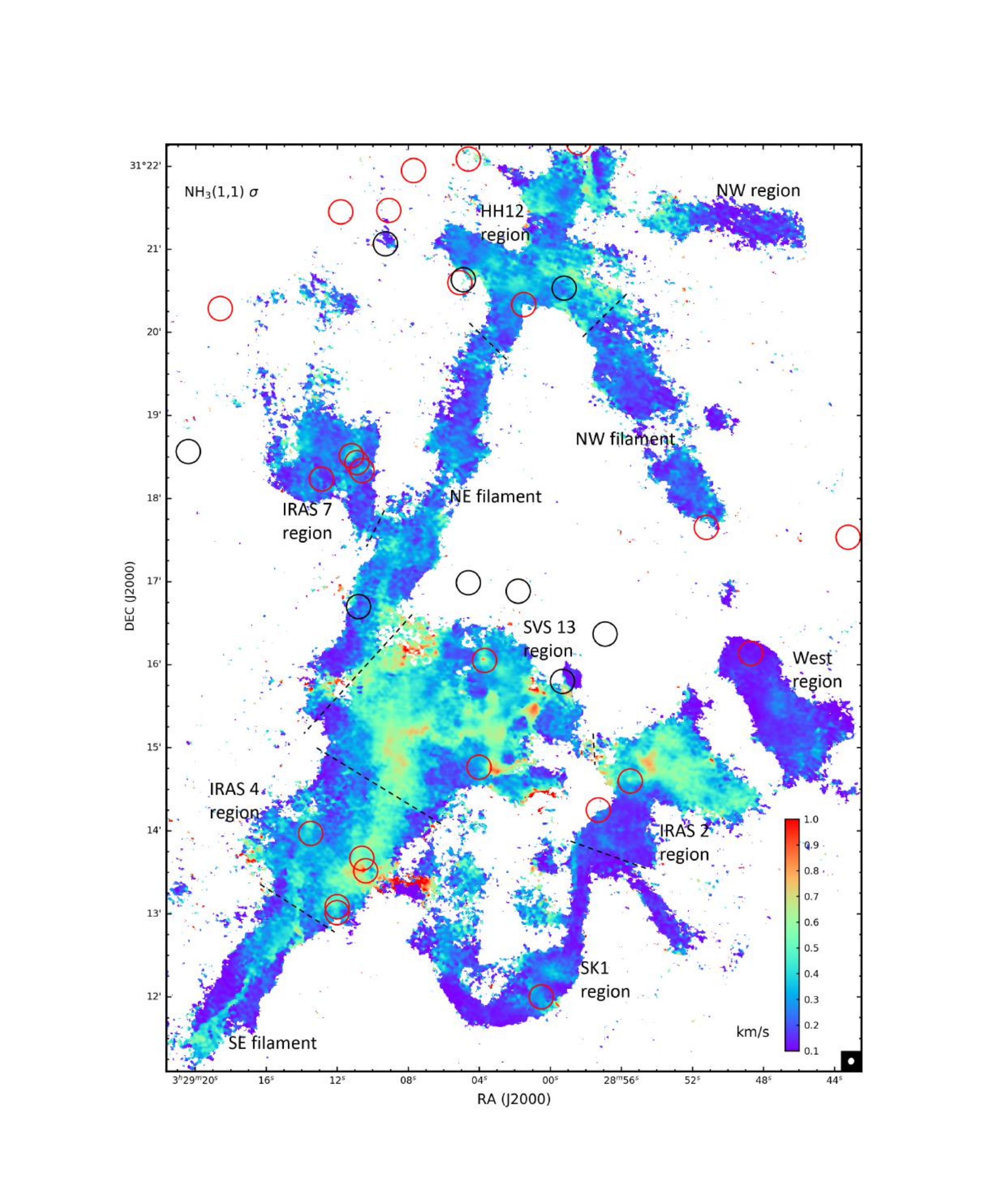}
\caption{NGC 1333 velocity dispersion map from spectral fitting of \amA. The typical uncertainty is 0.045~km/s.}
\label{fig:AmmDisp}
\end{figure*}

Spectral line fitting is carried out to obtain the \amm\ velocity structure of NGC 1333. Since we noted that the relative intensities of the satellites do not match theoretical predictions in many areas, we allow the peak intensities for each of the five \amA\ components to vary. The magnetic dipole based hyperfine splitting within each of the 5 groups also have spreads ranging between 11~kHz and 56~kHz. As some of them are resolvable by our channel width of 11.4~kHz, we fit the spectra to a model having all 18 hyperfine components. Even though we vary the relative intensities ($a_n$) of the five groups of hyperfine components ($n$ = 1,2,3,4,5), we assume that the relative intensities ($w_i$) of the components (denoted by $i$) within each of the 5 groups to be based on the theoretical values \citep[as given in Table 15 of][]{Magnum2015}. We assume a single line-of-sight velocity $v_{los}$ and a Gaussian line shape with a single dispersion value $\sigma_v$ for all the 18 components. Thus we fit the spectrum at each pixel as a function of velocity $S(v)$ using the following equation:
\begin{equation}
S(v) = \sum_{n=1}^{5} a_n \medop\sum_i w_i \exp {\bigg(-\frac{v-v_{los}-c\Delta f_i/f_0}{2{\sigma_v}^2}\bigg)} 
\end{equation}
where $c$ is the speed of light, $\Delta f_i$ is the frequency offset of the hyperfine component relative to the line center frequency $f_0$ = 23.6944955~GHz. We solve for $a_1$, $a_2$, $a_3$, $a_4$, $a_5$, $v_{los}$ and $\sigma_v$. We mask the pixels in which none of the channels have more than 4\,$\sigma$ detections (noise RMS $\sigma$ = 0.004~Jy/beam). Least-squares fits are obtained after providing reasonable limits to the parameter search spaces. Pixels for which the fit solution involves any of the parameters reaching the limits are considered to be bad fits and are excluded. The pixels containing emission from outflows are identified by their unusually wide Gaussian fits and are also omitted. The results are shown in Figures \ref{fig:MainAmp}, \ref{fig:SatAmp}, \ref{fig:AmmVel} and \ref{fig:AmmDisp}. The intensity map distributions can be considered to be representative of the brightness temperature distributions.

For all these maps, we calculate a typical uncertainty, which is reported in the respective figures. For the peak intensity maps and velocity maps, while least-squares fitting the spectra for each pixel, we get the variance of each parameter estimate by diagonalizing the corresponding covariance matrices. The median value of the variances of all the included pixels is reported as the uncertainty. For the integrated intensity maps, the uncertainty is derived from the combination of the peak intensity and velocity dispersion values and uncertainties.

The peak intensity map for the main \amA\ component has a median value of 30~mJy/beam. This value corresponds to a brightness temperature of 4.9~K. The maximum brightness temperature goes up to as much as 20~K near the IRAS 2 sources. The four satellite components $a_1$, $a_2$, $a_4$ and $a_5$ have median brightness temperatures of 2.4~K, 2.0~K, 2.1~K and 2.1~K, respectively. The ratios between the different components are not consistent throughout the map, but typically indicate an optical depth of greater than 1 for the main component, while the satellite components can be considered to be optically thin.

On comparing the main component intensity map (Figure \ref{fig:MainAmp}) to the integrated intensity map (Figure \ref{fig:AmmM0}), we find many differences in the maximum intensity regions. The integrated intensity map has greater intensities near the Class 0/I YSOs, but the peaks of the brightness temperatures are well distributed throughout, including the filaments and regions having low integrated emission. This difference is particularly evident in the south west quadrant of the map. The SK 1 region, the West region and the southern part of the IRAS 2 region all have greater brightness temperatures compared to the northern part of the IRAS 2 region. Yet, the IRAS 2 northern region has greater integrated intensities.

Correspondingly, we see in Figure \ref{fig:AmmDisp} that the velocity dispersions are greater in the regions containing the protostars (although they don't always peak at the YSO locations, but around them). They are particularly high near the boundaries of the SVS 13 region going up to as much as 0.8~km/s compared to the nominal values of about~0.3 km/s in the filaments. Some of the high dispersions correspond to known locations of outflows as well. 

In some regions, we see sharp boundaries in the dispersion map where areas having different estimates of dispersion are adjacent to each other. The sharp boundaries are caused by the presence of multiple velocity components of comparable intensity. In such cases, depending on the relative intensities, the spectral line fitting, which fits a single component, may select one of the components or fit both of them with a Gaussian having a greater width. These multiple velocity components also tend to be located near the protostellar sources. In addition, the SE filament has a distinct narrow region of higher dispersion as well. This feature is caused by the presence of the two parallel velocity coherent sub-filaments partially overlapping in the line-of-sight \citep{Dhabal2018}. In the narrow region, the two components, which have comparable intensity get fit by a Gaussian with a greater dispersion value. These multiple velocity component spectra also affect the velocity map and the intensity maps, but the effects are less drastic in those maps.  

The intensities of the satellite components also show variations, even though they have a good positive correlation of about $0.8-0.9$ to each other and to the main component. In the ideal case of no anomalies, the correlation is expected to be 1. The F$_1$=0-1 satellite component has greater mean intensities over most of the regions than the other satellites, while the F$_1$=2-1 and the F$_1$=1-0 satellites have relatively lesser mean intensities of all the satellites. The F$_1$=1-0 emission has lesser contrast between the strong emission regions and the surroundings. A mechanism to explain the deviations from the theoretical relative intensities of the satellites was suggested by \cite{Stutzki1985}. In their paper, the hyperfine anomalies are caused by selective trapping of infrared photons in the (J,K) = (2,1) to (1,1) transition. They are generally associated with maser activity and are postulated to be associated to star forming clumps of high density. A suggested indicator of this effect involves the ratio of the intensities of the outer satellites or the inner satellites. Typically $T_B$(F$_1$=1-0)/$T_B$(F$_1$=0-1) and $T_B$(F$_1$=2-1)/$T_B$(F$_1$=1-2) are both expected to be less than one in the selective trapping based anomalous regions \citep{Stutzki1984}.

We generated maps (not shown) of these two intensity ratios $O$ = $T_B$(F$_1$=1-0)/$T_B$(F$_1$=0-1) and $I$ = $T_B$(F$_1$=2-1)/$T_B$(F$_1$=1-2). The anomalous regions were defined to be ones where $O$ is less than 0.8 or greater than 1.2. By this definition, 55\% of the unmasked pixels are found to be anomalous. We found that the anomalous regions are spread throughout the map area, but are in lesser concentration in regions of high \amA\ integrated intensity. The lower ratio ($O<0.8$) regions comprise 67\% of the anomalous pixels, while the higher ratio regions ($O>1.2$) are concentrated mainly near the IRAS 4 region. Also we found no correlation between $O$ and $I$ (Spearman correlation coefficients between -0.1 and +0.1), even after selecting only the lower ratio anomalous pixels. Thus, the selective trapping mechanism fails to account for the relative intensity anomalies in the NGC 1333 region.

The line-of-sight velocity map shows rich variation throughout the map with values ranging from 6.25~km/s to 9.25~km/s. The velocity gradient in the SE filament \citep{Dhabal2018}, extends to the IRAS 4 region and further north to the SVS 13 region. The west part of the SVS 13 region has velocities varying monotonically by 2.0~km/s in the north-south direction. Similar large gradients are present in the northern part of the HH 12 region (in the east-west direction), and immediately to the west of the IRAS 4 sources. The IRAS 2 region also has a large magnitude velocity gradient in the SE-NW direction, but it is not monotonic. Most of these large gradients are caused by a combination of (a) multiple velocity-coherent components with varying intensities across the field of view and (b) a gradient across a single velocity-coherent component. 

The NE filament has a single velocity coherent component in most of the northern half and it has a gradient across it which changes direction once. Going from east to west, the gradient changes from positive to negative near the ridge. In the southern half, again there is a gradient across the filament which changes from negative in the east to positive in the west, but this seems to be caused by multiple velocity components in the line of sight. The velocity dispersion map of this region shows a sharp contrast region along the filament ridge. The NW filament, SK 1 region, the West region and the IRAS 2 have relatively lesser variations of about 0.5~km/s. The NW region has the least line-of-sight velocity variations among all fields. 

\subsection{Ammonia (2,2) Results}

\begin{figure*}
\centering
\includegraphics[trim={2.45cm 2.3cm 2.7cm 3.35cm},clip,width=\textwidth]{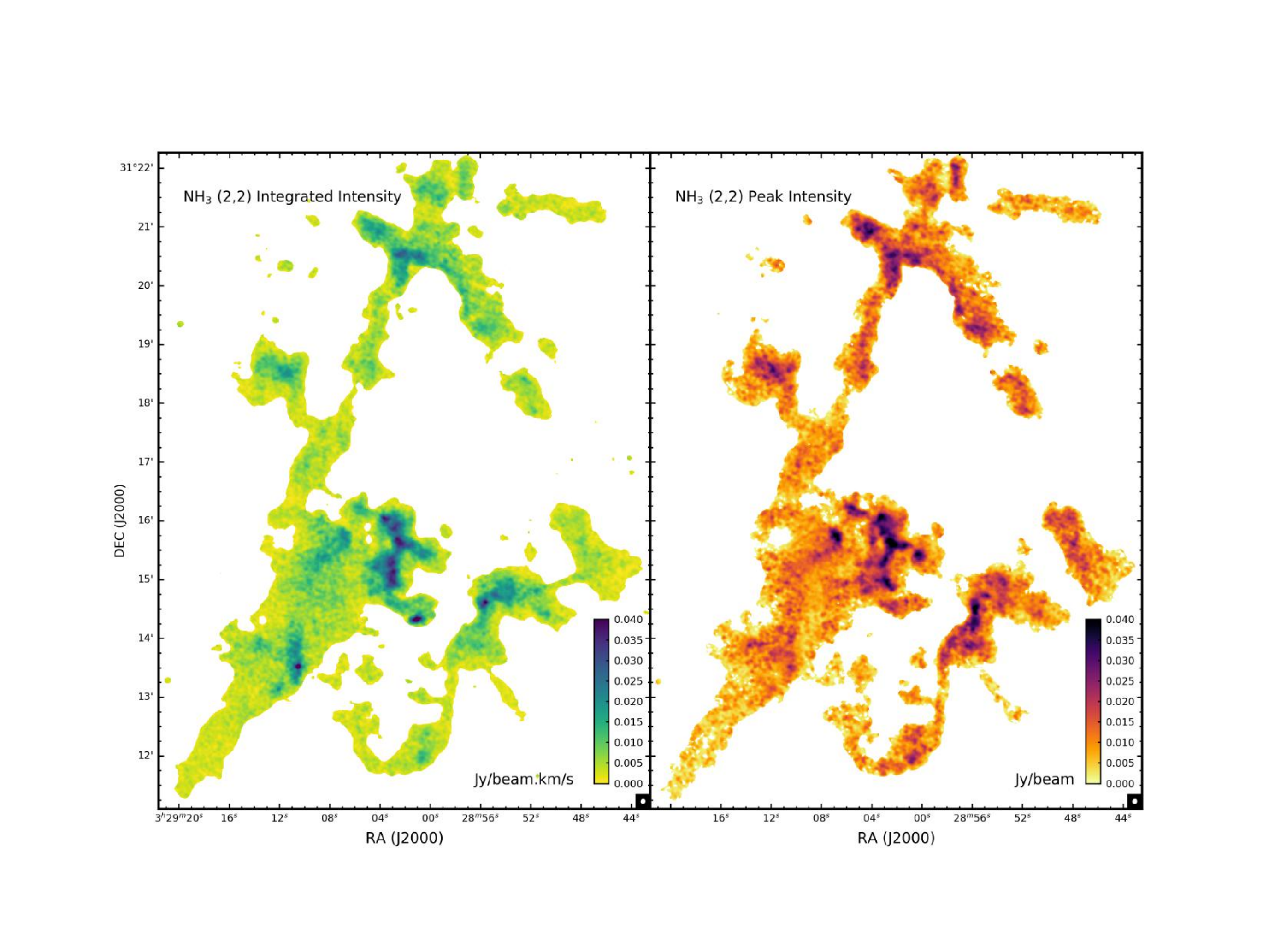}
\caption{\amB\ integrated intensity map (left) and \amB\ main component peak intensity map (right) of the NGC 1333 region. The typical uncertainties are 0.004~Jy/beam~km/s and 0.003~Jy/beam, respectively.}
\label{fig:Amm2Amp}
\end{figure*}

For the integrated intensity map of \amB\ (Figure \ref{fig:Amm2Amp}, left), we used a lower clip value of 2 times the RMS noise to cover a greater area of the map, that would be useful for the analyses in the following section. The \amB\ detections are spread over a relatively lesser area than (1,1) and they peak closer to the Class 0/I YSOs compared to the \amA\ integrated intensity emission. An outflow (from IRAS 2A) between the IRAS 2 region and the West region has \amB\ emission but is absent in the \amA\ map.

The satellite components of \amB\ have very low relative intensities and are indistinguishable from the noise for most regions (see Figure \ref{fig:SpectAmm}, bottom panel). We fit the central component of the \amB\ spectra with 12 hyperfine components, using their theoretical weight ratios. We fixed the line-of-sight velocities from the corresponding values of the \amA\ map to ensure fitting for the same velocity component as in \amA. In single component regions, solving for the line-of-sight velocity gives the same result in both \amm\ transitions, but this is not always followed in regions having multiple components in the line of sight. Hence, we solved for the velocity dispersion and a single peak intensity for the main \amB\ component. The peak intensity map is shown in Figure \ref{fig:Amm2Amp}, right. The median value is 9~mJy/beam corresponding to a brightness temperature of 1.5~K. The maximum brightness temperature is 7.5~K near the protostellar sources in the SVS 13 region. For the (2,2) transition, there is a better match between the integrated intensity map and the peak intensity map. Only the IRAS 2 region shows a reversal in the areas of maximum intensity for the two maps, i.e., in this region the areas having greater \amB\ integrated intensity have lesser \amB\ velocity dispersions and vice-versa.

\section{Analysis}
\label{sec:Ana}

\subsection{Optical Depth}

The optical depths can be estimated using the ratios of the main and satellite components of \amA\ assuming the hyperfine levels to be in local thermodynamic equilibrium (LTE). In this approximation, the optical depth ratios of the main to satellite components are equal to their respective theoretical intensity ratios ($\alpha$). The optical depth at any of the \amm\ inversion transitions' line centers $\tau_m$ can be calculated by solving for the equation \citep{Magnum2015}:
\begin{equation} \label{eqn:C2}
\frac{T_R(m)}{T_R(s)} = \frac{1-e^{-\tau_m}}{1-e^{-\alpha\tau_m}}
\end{equation}
where $T_R(m)$ and $T_R(s)$ are the main and satellite component source radiation temperatures. Their ratio is equal to the ratio of the respective relative intensities $a_m$ and $a_s$ obtained from the spectral line fitting. For either of the outer satellites of \amA, the theoretical value of $\alpha$ is 2/9. For the inner satellites, $\alpha$ is 5/18. For the NGC 1333 region, the LTE assumption is not accurate based on the anomalies presented in Section \ref{amm11}. So we investigated a modified LTE assumption using two satellite components having the same theoretical $\alpha$ value. In this method, we calculated the optical depth for the main component separately using the mean outer satellite intensities $\tau_{m,outer}$ and using the mean inner satellite intensities $\tau_{m,inner}$. The mean was taken to reduce the effect of the intensity anomalies, and to get a better optical depth estimate. $\tau_{m,outer}$ was found to be systematically higher than $\tau_{m,inner}$ based on the relative intensities. The variations in the intensity distribution get magnified in the $\tau_m$ maps and using the mean of the two inner or the two outer satellites does not nullify this effect. In some regions, the intensity ratios are so skewed, that we get non-physical values of $\tau_m$, particularly for $\tau_{m,inner}$. Due to these effects, over the entire map area, the $\tau_{m,inner}$ and $\tau_{m,outer}$ only have a modest positive correlation of about 0.31.

\begin{figure*}
\centering
\includegraphics[trim={2.5cm 2.0cm 2.9cm 3.0cm},clip,width=0.85\textwidth]{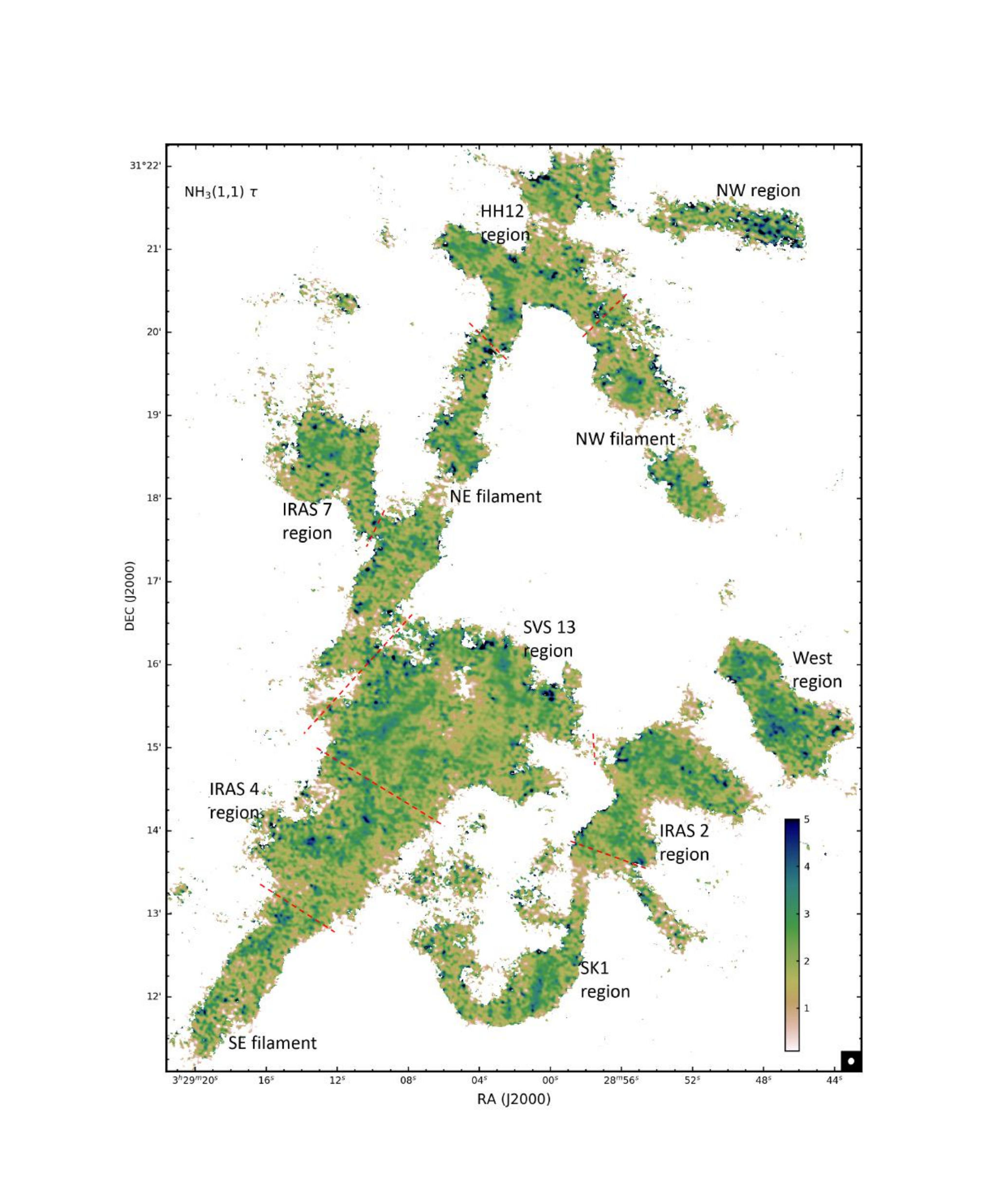}
\caption{NGC 1333 optical depth map at the \amA\ main component line center obtained using the ratio of all the satellite components to the main component of \amA\ (after appropriate scaling). The typical uncertainty in the optical depth values is 0.45.}
\label{fig:Ammtau}
\end{figure*}

Thus, the assumption involving two satellite components is only partially successful in deriving a consistent optical depth. Consequently, we add another modification to the LTE approximation by using all the four satellite intensities. To obtain a representative $\tau_{(1,1,m)}$ for \amA, we (i) scale up the outer satellite intensities by a factor of 5/4 (ratio of theoretical intensities), (ii) take the mean of all four intensities and (iii) solve equation \ref{eqn:C2} using a value of 5/18 for $\alpha$. The resulting optical depth map is shown in Figure \ref{fig:Ammtau}. All the regions are fairly optically thick (median value 2.1), with no particularly identifiable distribution pattern. We use this value of $\tau_{(1,1,m)}$ for the next analysis steps.

\subsection{Temperatures} \label{sec:Temp}

The level populations of the two quantum states ($n_+$ and $n_-$) involved in the \amA\ inversion transition are related by an excitation temperature $T_{ex}$ in the equation 
\begin{equation}
\label{eqn:T1}
\frac{n_+}{n_-} = \frac{g_+}{g_-} e^{-h\nu/kT_{ex}}
\end{equation}
where $g_+$ and $g_-$ are the statistical weights of the two levels. The statistical weight ratio is 1 for the inversion transitions of \amm. The transition frequency is $\nu$. The Boltzmann constant and the Planck constant are denoted by $k$ and $h$, respectively. The sum of $n_+$ and $n_-$ gives the total level population of the corresponding J=K quantum state.

\begin{figure*}
\centering
\includegraphics[trim={2.5cm 2.0cm 2.9cm 3.0cm},clip,width=0.85\textwidth]{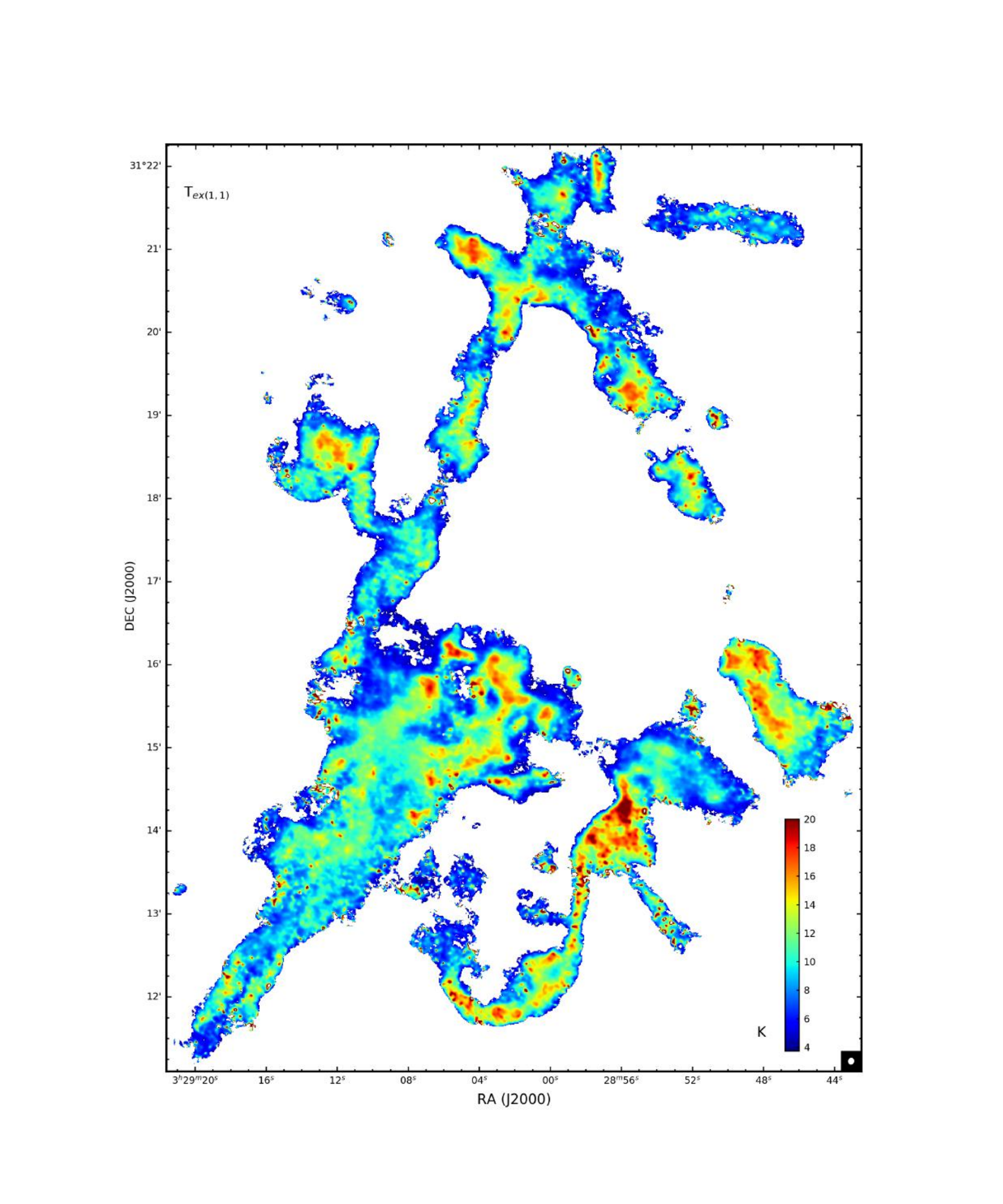}
\caption{\amA\ excitation temperature map of NGC 1333. The typical uncertainty is 0.6~K.}
\label{fig:AmmTex}
\end{figure*}

If we assume that a single value of $T_{ex}$ is applicable to all the hyperfine components of the inversion transition, we can use the optical depth $\tau_{(1,1,m)}$ and the brightness temperature $T_{b(1,1,m)}$ of the main component of \amA\ to determine the \amA\ excitation temperatures $T_{ex(1,1)}$ by solving for the equation
\begin{equation}
\label{eqn:T2}
T_{b(1,1,m)} = \Big(\frac{1}{\mathcal{J}_{\nu}(T_{ex})-\mathcal{J}_{\nu}(T_{bg})}\Big)[1-e^{-\tau_{(1,1,m)}}]
\end{equation}
where we have assumed a beam filling factor of 1. The (1,1) inversion transition frequency is $\nu$ and $\mathcal{J}_{\nu}(T)$ is the Rayleigh-Jeans equivalent temperature at a frequency $\nu$ of a black body at temperature $T$. It is given by $\mathcal{J}_{\nu}(T)$ = $\frac{h\nu}{k}/[\exp(\frac{h\nu}{kT})-1]$. The background radiation temperature $T_{bg}$ is 2.73~K. The resulting excitation temperature map is shown in Figure \ref{fig:AmmTex}. The median $T_{ex(1,1)}$ value is 9.4~K and it increases to a maximum of 20~K in the IRAS 2 region.

Similar to $T_{ex}$, a rotational temperature $T_{rot}$ can be used to define the ratio of the \amm\ level populations for (J,K) = (1,1) and (2,2) (column densities $N_{(1,1)}$ and $N_{(2,2)}$, respectively):
\begin{equation}
\frac{N_{(2,2)}}{N_{(1,1)}} = \frac{5}{3} e^{-41.0/T_{rot}}
\end{equation}

The factor 5/3 comes from the J-degeneracies and 41.0~K corresponds to the energy difference between the (1,1) and the (2,2) levels. From this equation, it is possible to derive $T_{rot}$ in terms of the \amA\ optical depth $\tau_{(1,1,m)}$, the ratio of the \amA\ and \amB\ main component intensities ($T_{R(2,2,m)}/T_{R(1,1,m)}$) and the ratio of the \amA\ and \amB\ velocity dispersions ($\sigma_{v(2,2)}/\sigma_{v(1,1)}$) as follows \citep{Magnum1992}:
\begin{equation} 
T_{rot} = -41.0 \bigg\{\ln{\Big[-\frac{0.283 \sigma_{v(2,2)}}{\tau_{(1,1,m)} \sigma_{v(1,1)}} \ln{\Big( 1-\frac{T_{R(2,2,m)}}{T_{R(1,1,m)}}(1-e^{-\tau_{(1,1,m)}})\Big)} \Big]} \bigg\} 
\end{equation}

The rotational temperature can be used to solve for the kinetic temperature $T_K$ in the following equation \citep{Swift2005} based on statistical equilibrium and detailed balance
\begin{equation} 
T_{rot} = T_{K}\Big\{ 1 + \frac{T_K}{41.5} \ln \big(1 + 0.6 e^{-15.7/T_K} \big) \Big\} 
\end{equation}

\begin{figure*}
\centering
\includegraphics[trim={2.5cm 2.0cm 2.9cm 3.0cm},clip,width=0.85\textwidth]{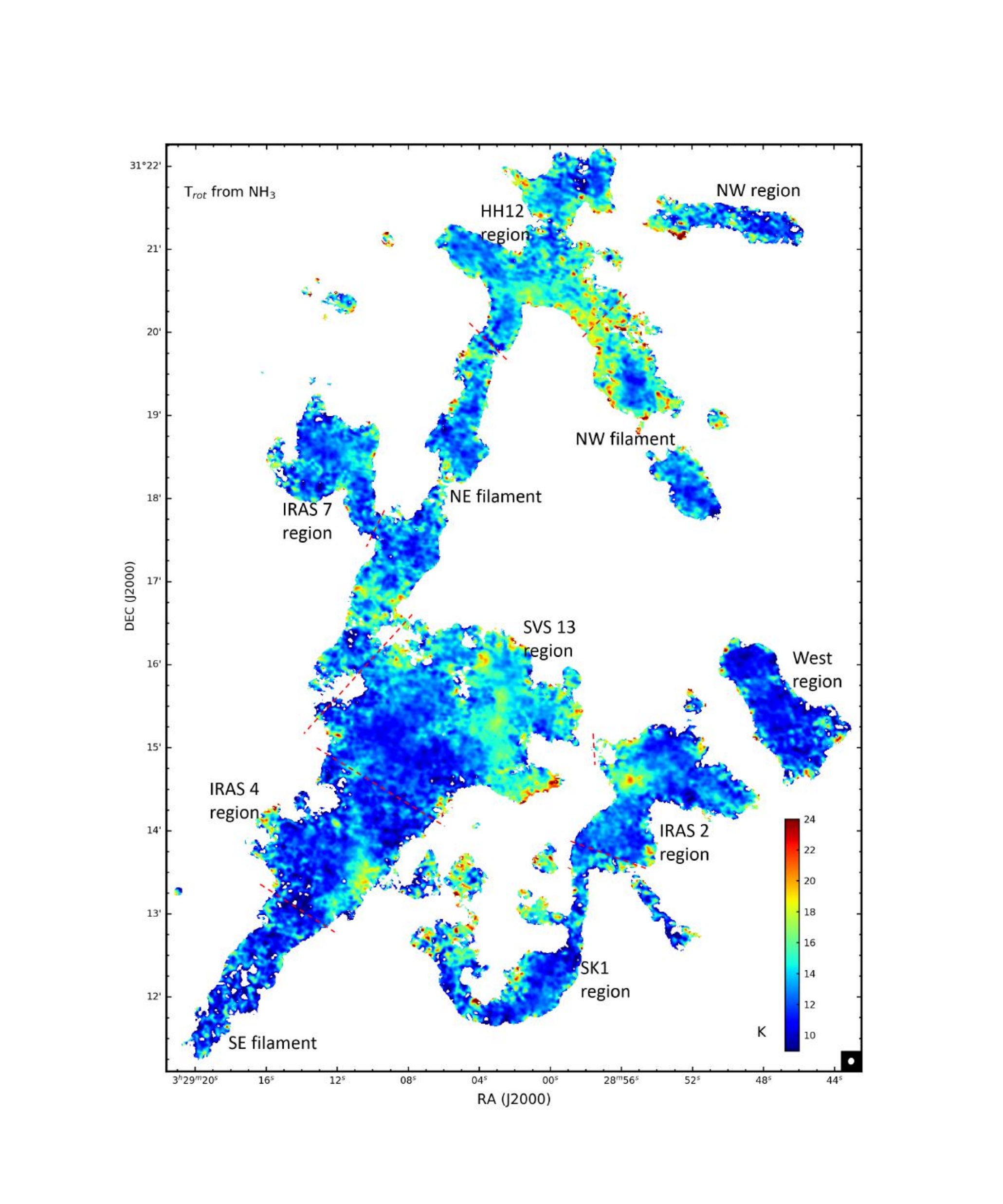}
\caption{NGC 1333 rotational temperature map obtained from the \amA\ and \amB\ maps. The typical uncertainty is 2.1~K.}
\label{fig:AmmTrot}
\end{figure*}

\begin{figure*}
\centering
\includegraphics[trim={2.5cm 2.0cm 2.9cm 3.0cm},clip,width=0.85\textwidth]{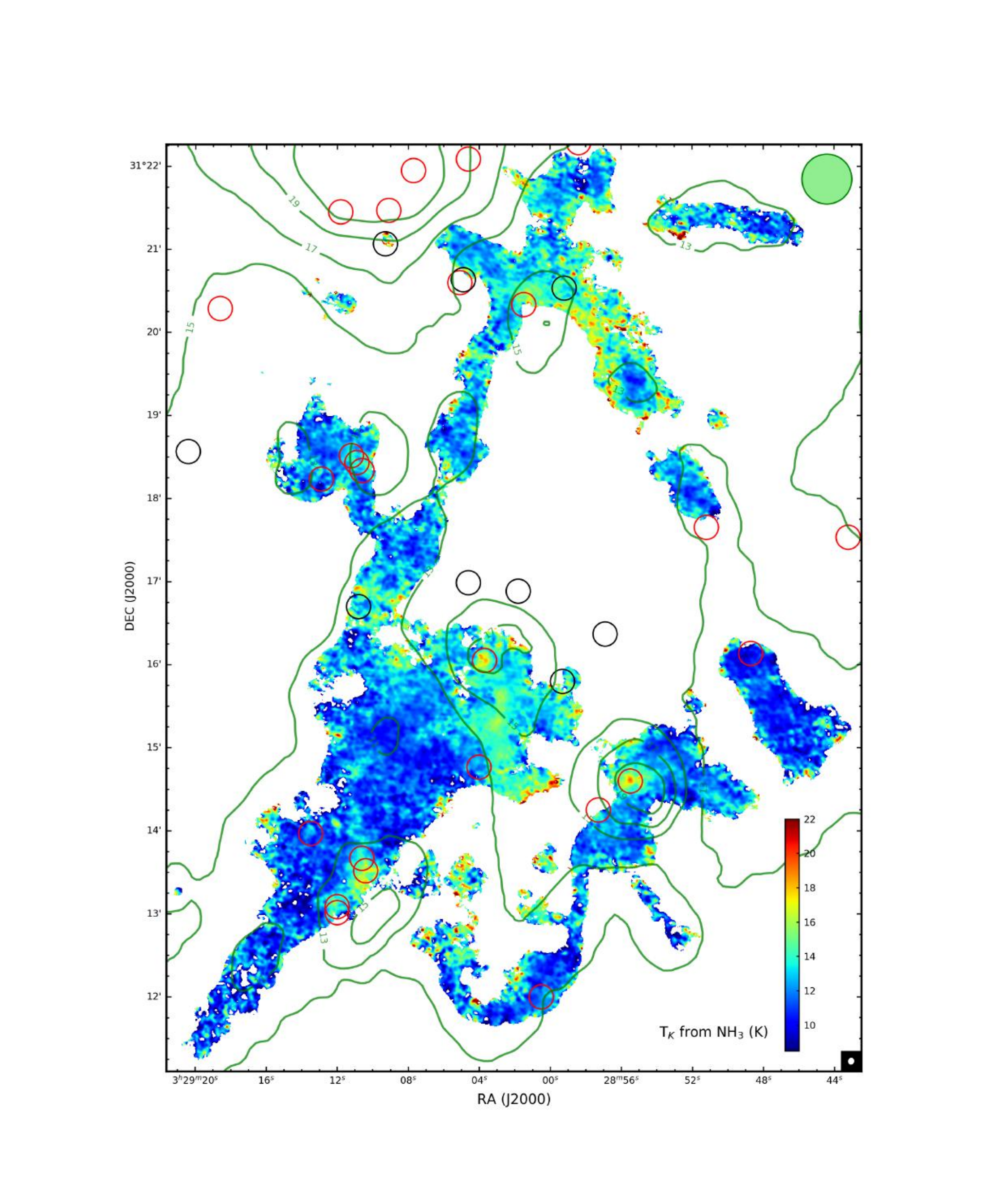}
\caption{NGC 1333 kinetic temperature map obtained from the rotational temperature map. The typical uncertainty is 1.8~K. The Class 0/I/Flat YSOs are overlaid as in Figure \ref{fig:AmmM0}. The contours of the \textit{Herschel} temperature map (in K) are overlaid \citep{Pezzuto2017}. The contour levels are 11~K, 13~K, 15~K, 17~K, 19~K and 21~K. The \textit{Herschel} beam is shown at the top-right corner.}  
\label{fig:AmmTK}
\end{figure*}

The rotational and kinetic temperature maps are shown in Figures \ref{fig:AmmTrot} and \ref{fig:AmmTK}, respectively. Both maps have very similar distributions. The median values of $T_{rot}$ and $T_K$ are 13.0~K and 12.4~K, respectively. In central NGC 1333, there are many regions of high temperature near the three main Class 0/I YSO groups. In addition, the west part of the SVS 13 region has higher temperatures between 14~K and 17~K. At the southern-west tip of this region, the temperature reaches 20~K. The northern part of the NW filament near the HH 12 region also has higher temperatures of $16-18$~K compared to the other filaments. 

Figure \ref{fig:AmmTK} also has the \textit{Herschel} dust temperature map contours \citep{Pezzuto2017} overlaid for comparison. Most of the filaments have dust temperatures between 11~K and 13~K, less than the temperatures ($\sim$~14~K) in the less dense areas of the cloud. The dust temperature values in the filaments and quiescent regions are in fair agreement to the kinetic temperatures obtained for \amm, indicating that the dust is thermodynamically coupled to the dense gas \citep{Forbrich2014}. The dust temperatures,  however, are greater than the \amm\ temperatures near the YSOs, with values greater than 20~K near IRAS 4, IRAS 2 and SVS 13. The dust temperatures are also greater than 16~K near IRAS 7 and HH 12. In these regions, the effective dust temperatures are elevated because the line-of-sight dust emission is dominated by contributions from the warm regions near the protostars, unlike the \amm\ which traces the entire column of dense gas \citep{Schnee2006}. We also note that in regions closer to the YSOs, there are greater temperature variations in the high resolution \amm\ map than in the dust map.

\subsection{Column Density}

To derive the column density of \amm\ we first express the column density of the upper level of the (1,1) inversion transition $N_{(1,1),+}$ in terms of the optical depth $\tau_{(1,1)}$ and the excitation temperature $T_{ex(1,1)}$ using the following equation
\begin{equation}
\label{eqn:CD1}
N_{(1,1),+} = \frac{3h J(J+1)}{8\pi^3\mu^2 K^2} \frac{1}{(e^{h\nu/kT_{ex(1,1)}}-1)} \int \tau_{(1,1)}dv
\end{equation}
where the dipole moment of the \amm\ molecule $\mu$ has a value of 1.468 Debye.
Using equation \ref{eqn:T1}, the ratio of the column densities of the two inversion states in (1,1) can be written as 
\begin{equation}
\frac{N_{(1,1),+}}{N_{(1,1),-}} = e^{-h\nu/kT_{ex(1,1)}}
\end{equation}
where $N_{(1,1),-}$ is the column density of the lower energy level of \amA. Hence, the total column density of the \amA\ is 
\begin{equation}
\label{eqn:CD2}
N_{(1,1)} = N_{(1,1),-} + N_{(1,1),+} = N_{(1,1),+}(e^{h\nu/kT_{ex(1,1)}}+1)
\end{equation}

Assuming a single rotational temperature $T_{rot}$ to define all the (J,K) level populations, the total \amm\ column density $N_{tot}$ can be estimated from $N_{(1,1)}$ and the rotational partition function $Q_{rot}$ using the following equation \citep{Magnum1992}:
\begin{equation}
\label{eqn:CD3}
N_{tot} = \frac{N_{(1,1)} Q_{rot}}{g_Jg_Kg_I}e^{E_{(1,1),+}/kT_{rot}} 
\end{equation}
where $g_J$ and $g_K$ are the rotational degeneracies and $g_I$ is the spin degeneracy. For \amA\ their values are 3, 2 and 0.25, respectively. $E_{(1,1),+}$ is the energy of the upper level in the \amA\ inversion transition. The value of $E_{(1,1),+}/k$ is 24.35~K where $k$ is the Boltzmann constant. This derivation uses a partition function that includes both ortho and para species of \amm.

By combining equations \ref{eqn:CD1}, \ref{eqn:CD2} and \ref{eqn:CD3}, we get
\begin{equation}
N_{tot} = \frac{3h J(J+1)}{8\pi^3\mu^2 K^2} \frac{Q_{rot}}{g_Jg_Kg_I} e^{E_{(1,1),+}/kT_{rot}} \frac{e^{h\nu/kT_{ex(1,1)}}+1}{e^{h\nu/kT_{ex(1,1)}}-1} \int \tau_{(1,1)}dv
\end{equation}
Since the optical depths for the different components of \amA\ are different, we express the total optical depth in terms of the main component optical depth of \amA\ using a relative intensity factor $R_i$ = 0.5. By using equation \ref{eqn:T2}, we can express this optical depth in terms of the integrated intensity of the main component of \amA\ ($\int T_{R(1,1,m)}dv$). We use a correction factor of $\tau/(1-e^{-\tau})$ for the optically thick case \citep{Goldsmith1999} which is applicable for most of the regions being analyzed. The resulting column density equation is thus: 
\begin{equation}
N_{tot} = \frac{3h J(J+1)}{8\pi^3\mu^2 K^2} \frac{Q_{rot}}{g_Jg_Kg_I} e^{E_{(1,1),+}/kT_{rot}} \bigg[\frac{e^{h\nu/kT_{ex(1,1)}}+1}{e^{h\nu/kT_{ex(1,1)}}-1}\bigg] \frac{1}{R_i} \bigg[\frac{\int T_{R(1,1,m)}dv}{\mathcal{J}_{\nu}(T_{ex})-\mathcal{J}_{\nu}(T_{bg})}\bigg] \frac{\tau_{(1,1,m)}}{1-e^{-\tau_{(1,1,m)}}} 
\end{equation}
The expression can be simplified to:
\begin{equation}
N_{tot} = 4.51 \times 10^{12}\ Q_{rot}e^{24.35/T_{rot}} \frac{1+ e^{1.137/T_{ex(1,1)}}}{1.517 - e^{1.137/T_{ex(1,1)}}} \frac{\tau_{(1,1,m)}}{1-e^{-\tau_{(1,1,m)}}} \int T_{R(1,1,m)}dv\ cm^{-2}
\end{equation}
where $\int T_{R(1,1,m)}dv$ is in K~km/s. The optical depths, excitation temperatures and rotational temperatures were calculated in the previous sections. To calculate the partition function $Q_{rot}$, we use the formula:
\begin{equation}
Q_{rot} = \sum_J \sum_K (2J+1) g_K g_I e^{-E_{JK}/kT_{rot}}
\end{equation}
The K-degeneracy $g_K$ is 1 for K = 0 and 2 for K~$\neq$~0. The nuclear spin degeneracy value $g_I$ is 0.25 for K~=~3n (ortho species) and 0.5 for K~$\neq$~3n (para species) , where `n' is a non-negative integer. The energy levels $E_{JK}$ up to J=5 are available in Table 9 of \cite{Magnum2015} and are sufficient for our rotational energy temperature ranges.

\begin{figure*}
\centering
\includegraphics[trim={2.5cm 2.0cm 2.9cm 3.0cm},clip,width=0.85\textwidth]{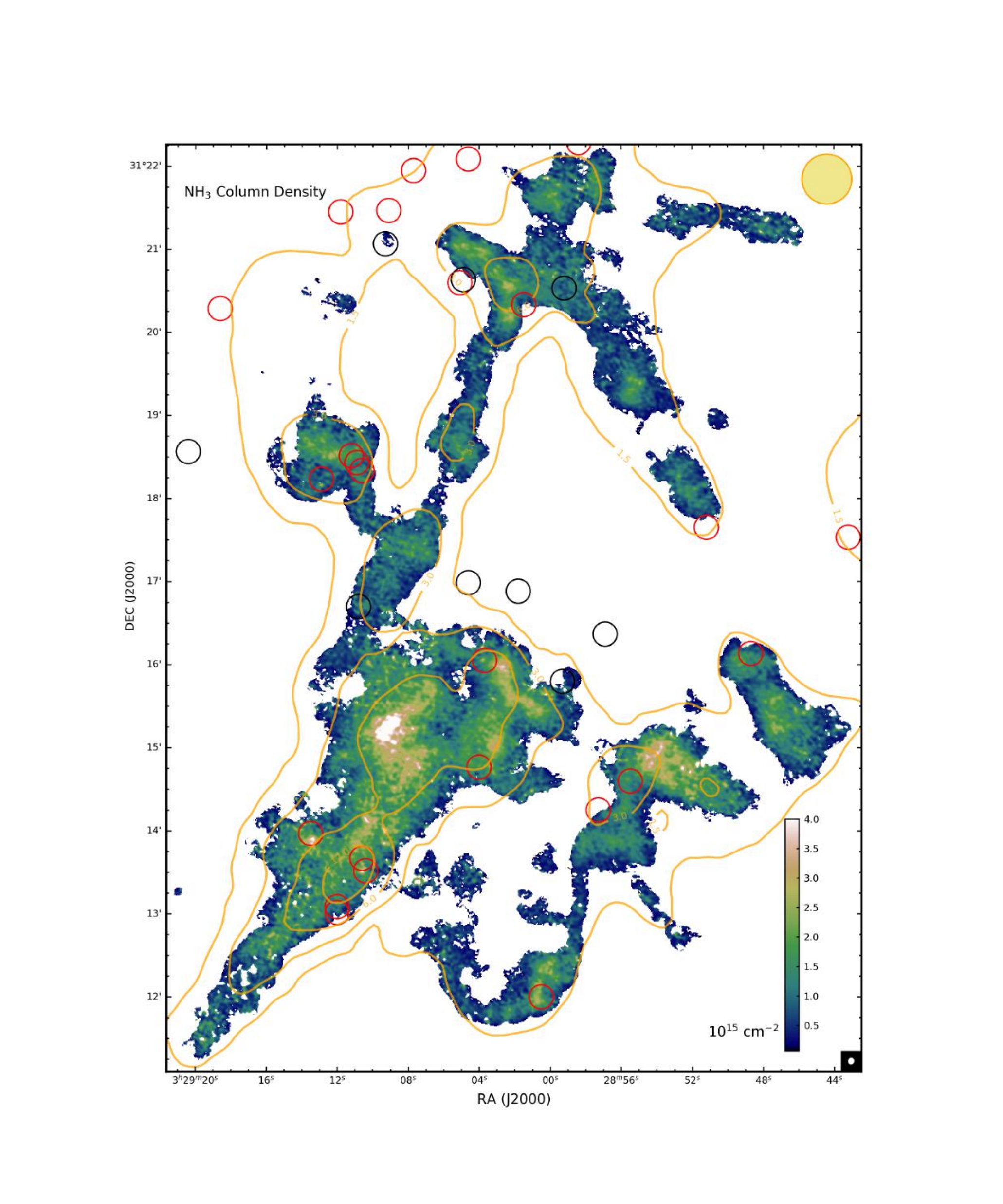}
\caption{\amm\ column density map for the NGC 1333 region. The typical uncertainty is 2.4~$\times$~10$^{14}$~cm$^{-2}$. The Class 0/I/Flat YSOs are overlaid as in Figure \ref{fig:AmmM0}. The contours of the \textit{Herschel} column density map are overlaid. The contour levels are 1.5, 3, 6 and 12 in units of 10$^{22}$~cm$^{-2}$. The \textit{Herschel} beam is shown at the top-right corner.}
\label{fig:AmmNtot}
\end{figure*}

The resulting column density map of \amm\ is shown in Figure \ref{fig:AmmNtot}. The column density values vary over two orders of magnitude, with a median value of 9.9~$\times$~10$^{14}$~cm$^{-2}$. In the eastern part of the SVS 13 region, the \amm\ column density is as high as 5~$\times$~10$^{15}$~cm$^{-2}$. This region coincides with the minimum dust temperatures ($\sim$~10.5~K) as well as gas kinetic temperatures ($\sim$~10~K) in NGC 1333. The filaments have lesser column density than the cloud center, with the NW filament having the least column densities (mean value of 7~$\times$~10$^{14}$~cm$^{-2}$). We overplot the dust-based H$_2$ column density map contours (from \textit{Herschel}) for comparison. The dust map traces the same structures as the \amm\ map (at the 36\arcsec\ \textit{Herschel} beam scale), including the narrow filament in the SK 1 region. The typical values are 3~$\times$~$10^{22}$~cm$^{-2}$ in the filaments and goes up to 9~$\times$~10$^{22}$~cm$^{-2}$ at the cloud cores. The only exception is the region close to the IRAS 4 sources, where the dust-based column density is much higher relative to the \amm\ column density in that region.

The expressions for the optical depth, the temperatures and the column density depend on the many terms and their error estimation is more involved. The previously determined uncertainties of the terms in the respective formulae are used as standard deviations of zero-centered normal distributions and added to the nominal pixel values of the terms to obtain alternative values. These alternative values are used to obtain alternative estimates of the expression being calculated at each pixel. The difference between the nominal expression and this alternative expression gives an error map. The standard deviation of the distribution of values in these error maps gives the respective uncertainties.

\section{Discussion}
\label{sec:Disc}

One important question about NGC 1333 that arises is how well the different \amm\ maps show the true characteristics of its dense gas and dust. In the previous section, we saw good spatial correlation with the \textit{Herschel} temperature and column density maps, but that comparison was limited by the 10 times larger beam size of Herschel-derived maps compared to our VLA and GBT combined maps. In this section, the regions are compared with the CLASSy-I \nnh\ maps covering the entire NGC 1333 region (Storm et. al. in prep.) and with the JCMT 850 \micron\ dust emission map \citep{Chen2016} from the JCMT Gould Belt Survey.

\subsection{Cloud Morphology}

\begin{figure*}
\centering
\includegraphics[trim={2.45cm 2.3cm 2.7cm 3.35cm},clip,width=\textwidth]{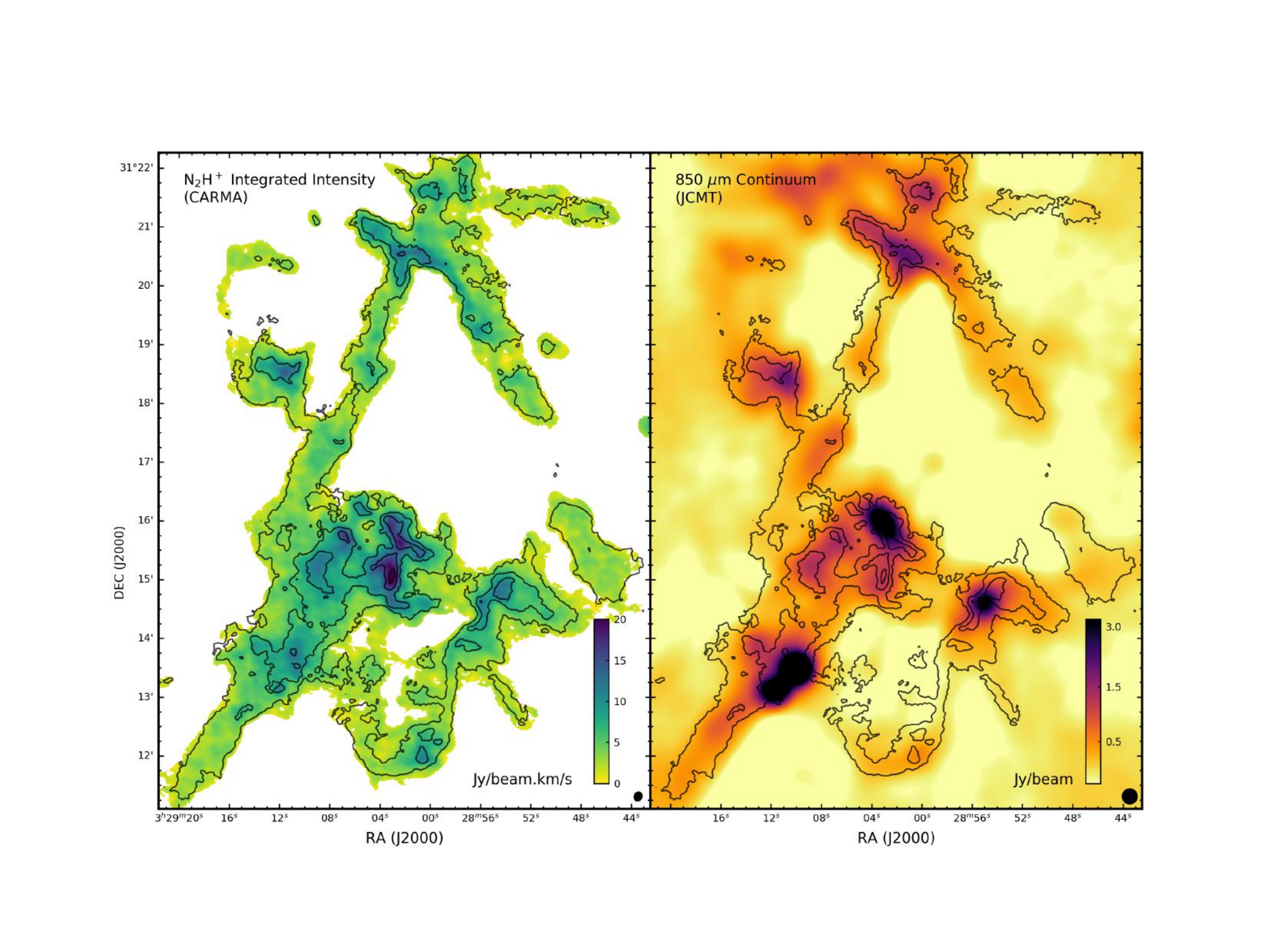}
\caption{\nnh\ integrated intensity map (left) and JCMT (SCUBA) 850 \micron\ map (right) of the NGC 1333 region with \amA\ integrated intensity contours overlaid on them. The contours are at 0.01, 0.02, 0.04 and 0.08~Jy/beam~km/s. The beams of the two maps are shown at the bottom right.}
\label{fig:M0Comp}
\end{figure*}

The CLASSy-I \nnh\ (J=1-0) maps at 93.173~GHz have a synthesized beam size of 9.06\arcsec~$\times$~7.58\arcsec, a channel width of 195.2~kHz (0.63~km/s) and a sensitivity of 0.08~Jy/beam \citep{Storm2014}. Figure \ref{fig:M0Comp} (left) shows the \nnh\ (J=1-0) integrated intensity map (color), with the \amA\ integrated intensity contours overlaid on it. Both integrated intensity maps include all the respective hyperfine components. The two maps show very good agreement on all scales. The \nnh\ maps have slightly better SNR which results in the emission having a greater extent, especially near the filaments. The only region of noticeable difference between the \nnh\ and \amA\ maps is at the north-eastern edge of the NW filament, which has relatively greater emission in \nnh. This location is the same place where higher temperatures were calculated from the \amA\ and \amB\ maps (see Section \ref{sec:Temp}).

The JCMT Gould Belt survey SCUBA observations included the NGC 1333 region \citep{COMPLETE2011}, with a map beam size of 15\arcsec\ at 850 \micron. A comparison with the \amm\ integrated intensity map (Figure \ref{fig:M0Comp}, right) also shows good correlation except in the regions associated with embedded YSOs, where the dust emission strongly increases due to increasing column density and temperature on small scales. On close comparison, we suspect that the JCMT map is offset by about half the beam size in the south-west direction. This offset is particularly evident in the NE filament, and the right arm of the narrow `u' shaped filament in the SK 1 region. One clear fact from the JCMT comparison is that the integrated intensity maps of \amm\ and \nnh\ are poor tracers of individual YSOs. This conclusion is also clear from the \amm\ column density map in Figure \ref{fig:AmmNtot}.

The high values in the \amm\ temperature map have good correspondence with the outflows of the region \citep{Plunkett2013} observed using CO maps and $H\alpha$ maps. The Herbig Haro objects HH 7-11 are associated with the high temperatures in the northern part of the SVS 13 region. The high temperature region in the HH 12 region and the connected NE filament is identified in the literature as a bow shock structure \citep{Knee2000}. The western part of the SVS 13 region and the parts of the IRAS 2 region are heated by the outflow from IRAS 2A.

\subsection{Cloud Kinematics}

\begin{figure*}
\centering
\includegraphics[width=\textwidth]{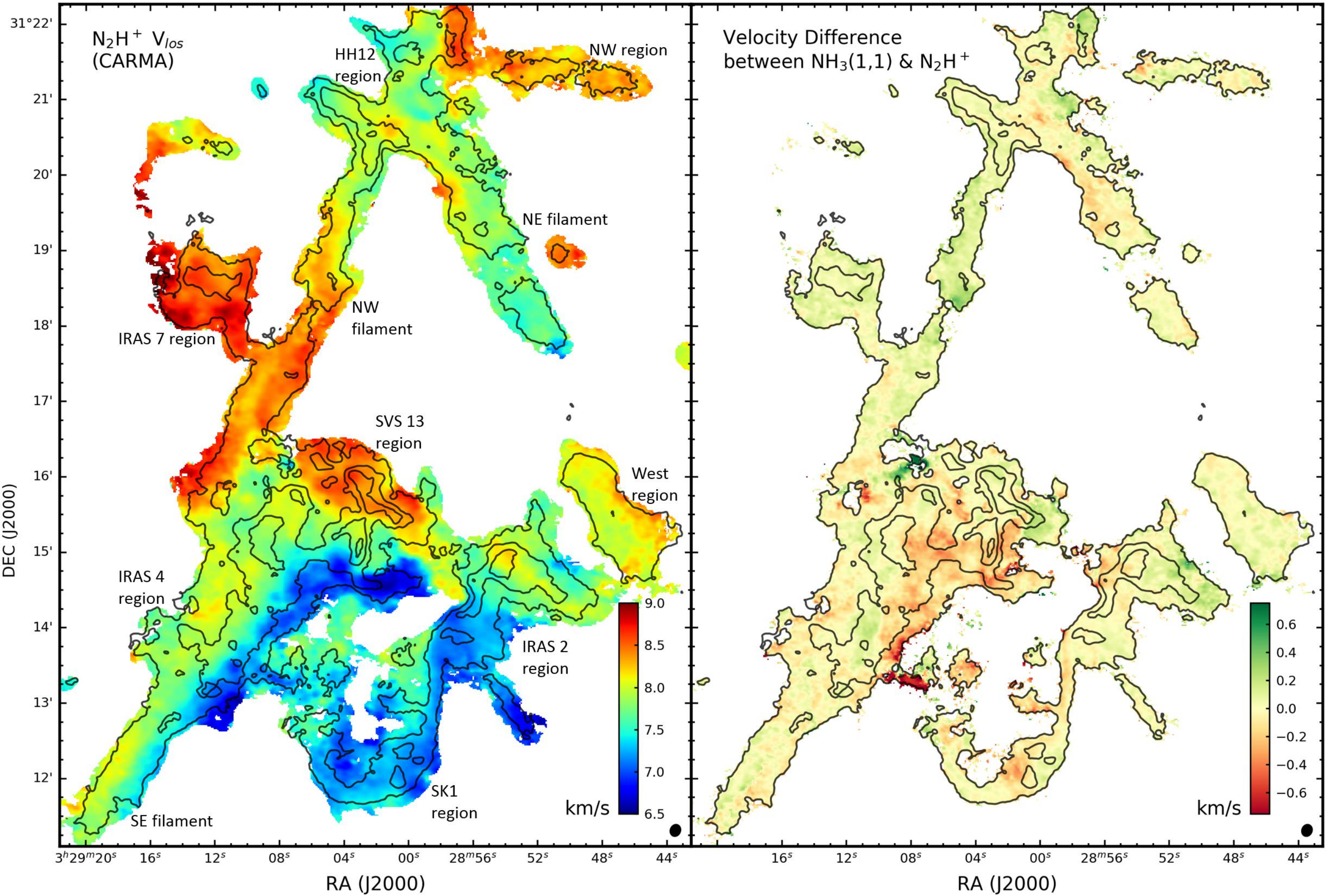}
\caption{\textit{Left:} \nnh\ line-of-sight velocity map of NGC 1333 obtained from spectral line fitting. The region names that are discussed in the paper are marked on this map. \textit{Right:} Difference between the \amm\ and \nnh\ velocity maps. \amA\ integrated intensity contours are overlaid on them as in Figure \ref{fig:M0Comp}.}
\label{fig:M1Comp}
\end{figure*}

To analyze the kinematics of \nnh\ relative to those of \amm, spectral fitting was carried out for all the seven hyperfine components of \nnh, with variable amplitudes for the three F$_1$ quantum number based groups of lines. The resulting line-of-sight velocity map and its difference with the convolved and regridded \amm\ velocities are shown in Figure \ref{fig:M1Comp}. The two maps are very well correlated in most regions to within the error in the \nnh\ spectral fitting (0.18~km/s). The results indicate that \amm\ is tracing the same material as \nnh\ and the dust, and that most features in the \amm\ map can be expected in high resolution maps of other similar dense gas tracers. 

Importantly, the velocity difference between an ionic species (\nnh) and a neutral species (\amm) can be used to infer ion-neutral drift, or ambipolar diffusion \citep{MS1956}. The general agreement in the velocity maps presented in Figure \ref{fig:M1Comp} therefore indicates that there is good ion-neutral coupling over most of the NGC 1333 region \citep{Flower2000}. This behavior is consistent with theoretical predictions that ambipolar diffusion is less efficient in quiescent environments \citep[see, e.g.,][for the typical timescale of ambipolar drift]{Spitzer1968,Mous1979}. 

Interestingly, the IRAS 4 and the SVS 13 regions show relatively large differences between the two velocities. These differences could be related to multiple velocity components as discussed in the following paragraph, but could not be ascertained using our data. Alternatively, the differences could also be related to large ion-neutral drift ($\gtrsim 0.2$~km/s), which could be explained by shock accelerations, as proposed by \citet{Li2004} and examined and confirmed in various numerical simulations \citep[e.g.,][]{Nakamura2005, Klessen2005, VS2011, ChenOst2012, ChenOst2014}. Considering these two regions potentially represent the shock front of an expanding bubble that is responsible for the formation of the SE filament (see discussions below in Section~\ref{sec:CollTur}), our results could be the first observational validation of shock-accelerated ambipolar diffusion. 

We note that there are two small regions with $\gtrsim 1$~km/s velocity difference between \nnh\ and \amm\ (the dark green area on the eastern side of SVS 13 and the dark red area on the western side of IRAS 4). In fact, on investigating our data with the F$_1$=1-0 isolated hyperfine component of \nnh\ (J=1-0), we found that multiple overlapping velocity-coherent components are commonly seen in those locations. In most cases, these multiple components have similar relative intensities for both molecular species, and thus the spectral fitting, which does a single velocity fit, gives comparable line-of-sight velocities. In the two small regions with $\gtrsim 1$~km/s velocity differences, however,
the relative intensities of the two velocity-coherent components in the two species are reversed, resulting in one getting selected by \nnh\ and another by \amm\ and hence the significantly larger velocity differences.

Since a lot of these regions have multiple velocity components, it is challenging to get a good estimate of the velocities of each of these components using complex molecular tracers like \nnh\ and \amm, given their many hyperfine components. Attempts to fit these spectra with two velocity components produced multiple solutions, that are very different from each other but have almost equal fitting errors. This issue is further compounded by the variation of the relative intensities of the hyperfine components from region to region. To get a good representation of all the multiple velocity components, some of these regions (especially near the Class 0/I YSOs) need to be observed with dense gas tracers like \htco. It is optically thin in most regions and has no hyperfine splitting, thereby allowing easier identification of the different velocity components. This tracer would also allow better estimation of the velocity dispersion of each component. A single component fit sometimes overestimates the dispersion in the dynamic environment near protostars having multiple velocity components.  

\subsection{Filaments in NGC 1333 and their relation to Star Formation}

We identified three filaments in the NGC 1333 region (the SE, NE and NW filaments) and some narrower filaments in the SK 1 region. The FWHM values of the filaments range from 0.015 pc in the SK 1 region to 0.05 pc in the NW filament. The corresponding deconvolved widths from the JCMT map after taking into account the larger beam size, show that the filament widths in dust are greater than those in dense gas by about 50\%. These results match those found for the CLASSy-II filaments \citep{Dhabal2018}.

\subsubsection{Filament formation by colliding turbulent cells}
\label{sec:CollTur}

\begin{figure*}
\centering
\includegraphics[width=0.6\textwidth]{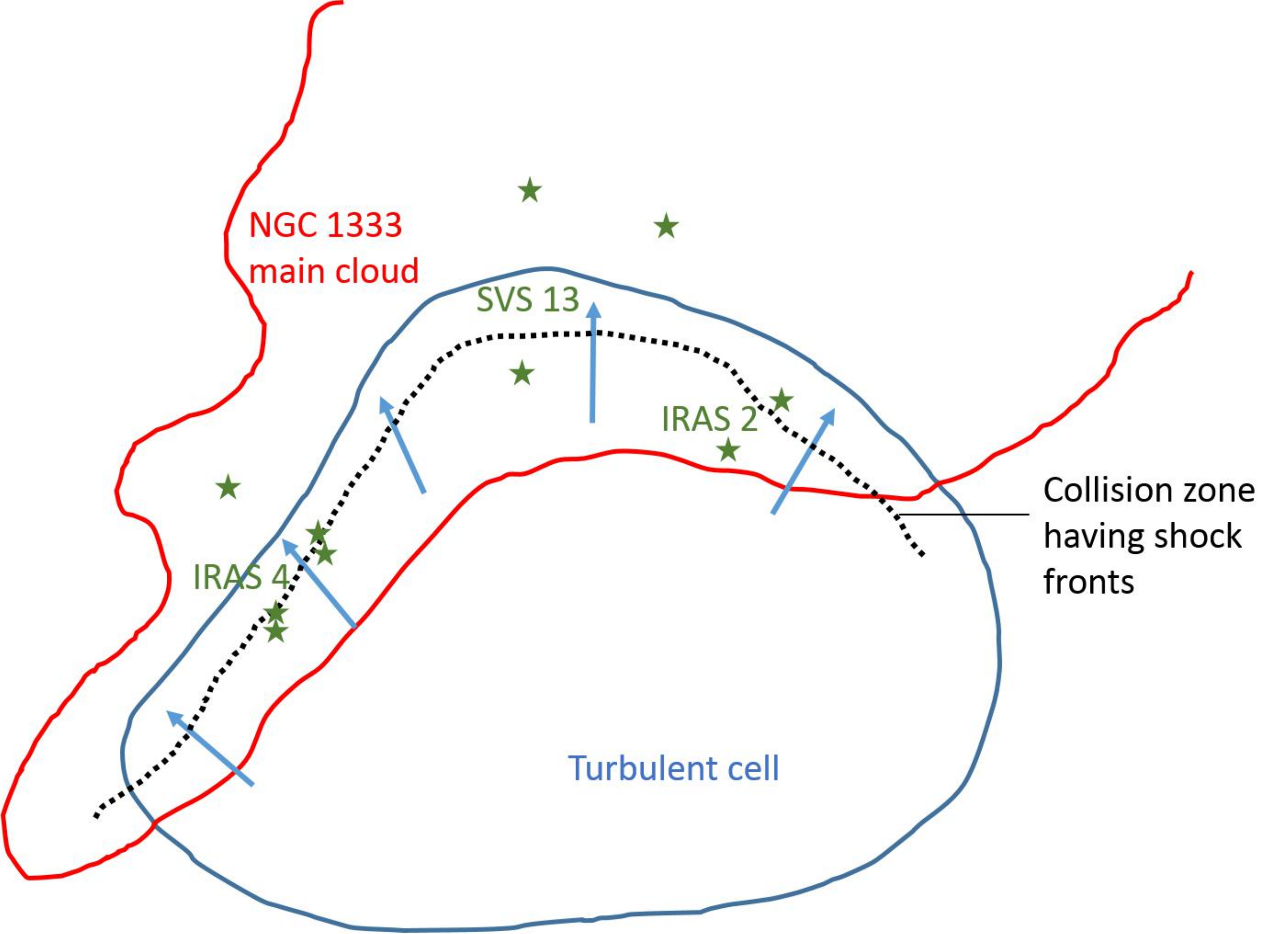}
\caption{Cartoon model of the southern half of the NGC 1333 region, illustrating the main cloud region (red boundary) and the proposed turbulent cell (blue boundary). The turbulent cell is farther from us in the line-of-sight than the NGC 1333 main cloud and is moving towards it. The dotted line marks the zone of collision, in which the IRAS 4, SVS 13 and IRAS 2 multiple protostellar systems are formed.}
\label{fig:Model}
\end{figure*}

The \amA\ map of the SE filament indicate a presence of two parallel sub-structures, as seen also in \htco\ and \nnh\ observations. The analysis of the \htco\ map of the same region showed that one of the two sub-structures has a significant velocity gradient across it \citep{Dhabal2018}. With our large-area, high-resolution \amA\ maps, we can put this in perspective of the kinematics of the entire cloud. We note that in the southern part of NGC 1333 there is a large region having line-of-sight velocities around $6-7$~km/s, which are at least 1~km/s less than those in adjacent regions. We see a large velocity gradient all along the southern edges of the SE filament, the IRAS 4 region, the SVS 13 region and the IRAS 2 region. The gradient is as much as 2~km/s across 0.09 pc in the SVS 13 region. It continues into the IRAS 4 region and the SE filament, where the gradients are comparable ($1-1.5$~km/s across 0.04~pc). The IRAS 2 region also shows a ridge of high velocity resulting in a sharp gradient in the SE-NW direction. We propose that this behavior could indicate a large-scale ($\sim$~0.5~pc) turbulent cell moving towards the cloud center from the south (in the projected sky plane). The collision between this cell with the cloud could result in the formation of a layer of compressed gas, from which dense star-forming clusters (IRAS 4 and IRAS 2) emerge, as well as an elongated structure (the SVS 13$-$IRAS 4$-$SE filament system) at the shock front. This scenario is illustrated in Figure~\ref{fig:Model}.

The importance of supersonic turbulence in structure formation is well established by numerical simulations with or without magnetic fields \citep{Padoan1999, Banerjee2006, PudKev2013}. Generally driven by stellar winds and supernovae \citep{MLKlessen2004}, these large-scale supersonic flows generate locally planar shock layers of compressed gas and help in transferring energy to smaller scales. In such shock layers, dense structures can form, either seeded by turbulent velocity perturbations \citep{ChenOst2014, ChenOst2015} or at the intersection of shock fronts \citep{PudKev2013}. With increasing density, gravity starts becoming dominant, amplifying the initial anisotropies and resulting in the formation of protostellar systems.

In the SE filament region, a velocity gradient was identified across the filament with \htco\ \citep{Dhabal2018}, which is also seen in \nnh\ (Figure~\ref{fig:M1Comp}) and \amm\ (Figure~\ref{fig:AmmVel}).
This feature is consistent with the filament formation model proposed by \cite{ChenOst2015} (see also Chen et al. {\it in prep.}). In this model, the velocity gradient is a projection effect of gravity-induced accretion within the shocked layer created by colliding turbulent cells. We note, however, that in this model the ratio between gas kinetic energy and gravitational energy, $C_v \equiv \Delta {v_h}^2 / (GM(r)/L)$, is typically $\sim~0.1$ from numerical simulations (Chen et al. {\it in prep.}). Here, $\Delta v_h$ is half of the velocity difference across the filament out to a transverse distance r from the filament spine, and $M(r)/L$ is the mass per unit length of the filament measured at the same distance from the spine as the $\Delta v_h$ measurement. For the SE filament, we estimated the mass per unit length based on the \textit{Herschel} column density map obtained from the Gould Belt Archive \citep{Sadavoy2014}, and derived $M/L \approx 20$~\Msun/pc \citep{Dhabal2018}. Combining with the measurement $\Delta v_h \approx 0.35$~km/s, we get $C_v~\sim~1.6$, much larger than those expected from gravity-induced velocity differences. This disparity suggests that the SE filament is at the front-end of the expanding bubble, and the velocity gradient observed across the filament is due to the shock, because gravity from the filament itself is not sufficient to generate that velocity difference.

There are additional observational evidences of the turbulent cell collision scenario. 
Firstly, previous studies on polarimetry around this area using optical/near-IR extinction \citep{Alves2011} and dust thermal emission (JCMT BISTRO; Doi et al. {\it in prep.}) 
have both found that the magnetic field within the SE filament is likely parallel to the filament, which is a feature of shock-compressed gas because shocks only amplify the magnetic field component that is parallel to the shock front \citep[see, e.g.,][]{Shu1992}. Secondly, the locations along the high velocity gradient ridge also have relatively high velocity dispersions ($>$ 0.5~km/s) in the \amA\ map that are typical for shocked layers. Thirdly, we see many protostars along the boundary of this proposed compressed gas layer. This region has the highest number of Class 0/I YSOs in NGC 1333. They are divided into the multiple systems IRAS 2, IRAS 4 and SVS 13.  The star formation in this region could be triggered by gravitational instabilities in the dense gas layer. Similar interpretations have been made for cloud-cloud collisions based on observations \citep{DC2011,Nakamura2014,Dewangan2017}, and such collisions are often reported in simulations \citep{VS2003, InFuk2013, Takahira2014}. Finally, water masers are identified near all three afore-mentioned protostellar systems along the velocity gradient ridge (see Figure \ref{fig:AmmM0}). The relative kinetic energies of shocked and unshocked gas are known to power masers; hence their presence is also an indication of shocks propagating in dense regions \citep{Elitzur1989, Hollenbach2013}.

We detect multiple parallel sub-structures in the SE and NE filaments, and all along the proposed compressed gas layer. Similar observations of parallel sub-structures in the Serpens filaments and in observations by other authors \citep{Beu2015, Hacar2013} indicate that such behavior is an important feature related to filament formation. It may be argued that this phenomenon is a density selection effect, wherein the filament spine has a density much greater than the critical density of the used tracers (Chen et al. {\it in prep.}). Thus, for these tracers, the gas in the filament spine emits less than the surrounding, less-dense part of the filament. Hence, a single physical filament could appear to be composed of multiple parallel sub-structures. This hypothesis may be tested using optically thin tracers of very dense gas such as \htcn\ emission. 

Considering the alternative that the multiple parallel structures are physically separate, a common formation mechanism like the `fray and gather' model \citep{Smith2016} has been proposed, based on hydrodynamic simulations of turbulent clouds. In this model, the sub-filaments are formed first and are then gathered together by large scale motions of the cloud. Alternatively, it could be explained as multiple filaments forming in the dense gas layer created by colliding turbulent cells, which on projection in the sky plane can seem parallel to each other. Another explanation involves formation of a wide filament that fragments into multiple sub-filaments \citep{Taf2015}.

The previously available GBT maps of NGC 1333 \citep{Friesen2017}, by themselves, were unable to resolve the rich structure within the filaments, all of which have narrower widths than the GBT beam size at 22~GHz. The red-shifted and blue-shifted parts of the cloud from which we propose the turbulent cell theory is also discernible in the GBT-only velocity map. The narrow shock zone with the high velocity gradient and high velocity dispersion, however, is resolved only in the combined VLA and GBT maps.

\subsubsection{Effect of Outflows on Filaments}

Outflows are known to clear the gas surrounding forming stars, and eventually lead to disruption of the larger cloud \citep{ArceThesis}. They could also gravitationally unbind the gas in the dense core and stop the stellar mass accretion phase \citep{Taf1997}. In the NGC 1333 region, there is more energy in outflows than is observed in bulk, turbulent motions. Thus despite having a tiny fraction of the total mass, outflows have a key role in shaping the region. It is estimated that at the current rate of energy ejection from the outflows in NGC 1333, the entire cloud will be dispersed in 10 million years \citep{Curtis2010}.

The NE filament and the NW filament are in locations known to be impacted by outflows. \cite{Knee2000} showed that the cavity between these filaments is filled with high-velocity gas from several outflows. They identified HH 12 as the leading bow shock of a Class 0 source in the SVS 13 multiple system. The region inbetween these filaments is clear of dust and gas, and many evolved YSOs and main sequence stars are present here. Past outflows from these sources could have played an important role in the formation of the NE and NW filament by clearing out the cloud. \cite{Quillen2005} interpreted the two filaments as the walls of ancient outflow cavities which are no longer actively driven. 

The NE filament shows major velocity variations along its length ($\sim$~0.7~km/s) and across its width ($\sim$~0.5~km/s) with a ridge of maxima in the north and a ridge of minima in the south. In addition to the evolved sources on the west of the NE filament, the outflow from one of the IRAS 7 sources \citep{Bally1996} can also contribute to the morphology of the NE filament from the eastern side. Based on the CO outflows identified by \cite{Plunkett2013}, we find that one of their outflow candidates `C3' is aligned perfectly along part of the eastern wall of the NE filament. 

The narrow short filament extending towards the south-west from the IRAS 2 region is also found to be parallel to the southern lobes of the outflows from IRAS 2A and IRAS 2B \citep{Plunkett2013}. In addition, the filament is coincident with the overlap region of these two outflows lobes. Although these evidences suggest that the short filament may have been formed in the shock front produced by the swept up gas between these two parallel outflows, this filament has relatively low \amA\ velocity dispersion values ($\sim$~0.16~km/s) in Figure \ref{fig:AmmDisp}. This is not expected for shocked regions. The narrow `u' shaped filament in the SK 1 region has also been proposed to be a dust shell by \cite{Lefloch1998}. However, as with the other filament in this region, this filament has low velocity dispersion values, which is not expected in bow shocks. The cavity to its north could still have been cleared out by outflows from one or more of the sources in the cloud core. Similarly, the IRAS 2 outflows may be responsible for clearing the material on the two sides of the narrow south-west filament. 

\subsubsection{Star Formation}

We find that many of the Class 0/I YSOs are embedded in the cloud, but not necessarily confined to the filaments. Of the isolated protostars, there is one Class 0/I source at the southern tip of the NW filament, and another similar source SK 1 in the `u' shaped filament. Most of the remaining Class 0/I sources are present in protostellar multiple systems like IRAS 2, IRAS 4 and IRAS 7. Of these, seven of the sources are present along the proposed high-velocity-gradient ridge identified in Section \ref{sec:CollTur}. The compression of the gas to very high densities can make that gas gravitationally supercritical and induce cores and protostars to form within them \citep{Balfour2015, Smith2014}. \cite{Belloche2006} came to similar conclusions about the origin of IRAS 4A and IRAS 4B based on the high mass infall rate measured in the IRAS 4A envelope, which is only possible in a model involving a strong external compression wave. 

\cite{Volgenau2006} used high resolution maps of the IRAS 2 and IRAS 4 protostellar systems to show that the cores harboring these sources are not quiescent homologous structures. Even on the scales of the protostellar envelopes, turbulence is the major contributor to the observed line-widths. Many of the sources such as IRAS 4A, IRAS 4B and IRAS 2A are close binary protostars. (Of these, IRAS 2A is not resolved in \textit{Spitzer} data; we mark it as a single protostar although it has been confirmed to be a binary using radio data.) Simulations have also shown that fast compression is often associated with core fragmentation \citep{Henn2003} which could explain the recent results reported by \cite{SS2017} that most stars form initially in wide binaries. In addition, based on the uncorrelated direction of the outflows from the binary system of IRAS 2A, \cite{Tobin2015} concluded that it is evidence of core/envelope fragmentation by turbulence and not disk fragmentation due to gravitational instability. These observations indicate that turbulence at different scales plays an important role in the star formation process, particularly in the southern half of NGC 1333.

There is a string of Flat YSOs in the less dense area north of the SVS 13 region. Along the NW filament, there is a sequence of five Class II YSOs. The relatively evolved YSOs are spread throughout the cloud including the lower density regions. Indeed, outflows from these more-evolved objects may have previously cleared out some regions of the cloud. The formation of the Class 0 YSO sources -- VLA 42 in the HH 12 region and SK 1 have been proposed to be triggered by outflow bow shocks by \cite{Sandell2001}. We do not find sufficient evidence for such shocks in the case of SK 1, because of the low velocity dispersion values along the proposed shock layer. In the case of VLA 42, however, there is evidence of high temperatures ($>$~16~K) and large velocity dispersions ($\sim$~0.4~km/s) in its environs over a large area (compared to the VLA beam area). Bright $H\alpha$ emission is also detected in this region \citep{Walawender2005}, which further supports a shock-triggered star formation scenario for VLA 42. Thus, outflows could be instrumental in regulating star formation in NGC 1333. 

\section{Summary}
\label{sec:Summ}

In this paper, \amm\ observations are used to study the morphology, kinematics and temperatures of the entire NGC 1333 region at an angular resolution of 4\arcsec. We showed the importance of high-resolution, large-area maps that are sensitive to structure at all the spatial scales. To attain coverage, VLA interferometric data were combined with GBT single dish maps in the visibility domain. We have studied the effects of the various parameters that are important in the joint deconvolution of the visibility data. 

For further analysis, we used data cubes corresponding to \amA\ and \amB\ inversion transitions, which have the brightest emission among all the observed transitions. The final data cubes have a channel width of 11.4~kHz, a synthesized beam of 3.93\arcsec~$\times$~3.41\arcsec, and a sensitivity of 4~mJy/beam. We produced the integrated intensity maps for both \amA\ and \amB. Line fitting was carried out taking into account all the hyperfine components of \amA\ to obtain the peak intensity maps of the main and satellite components, the line-of-sight velocity map and the dispersion map. For \amB, we used a similar method on the main component to obtain its peak intensity map and the dispersion map. We also derived maps of the optical depth for the \amA\ main component, the rotational temperature, the kinetic temperature and the column density of \amm.

Using these maps, we studied the distribution and properties of the dense gas in NGC 1333. Our main conclusions are summarized below.

\begin{itemize}

\item In locations near the cloud center and in some filaments, multiple velocity components in the cloud are observed. 

\item In many locations, there are large deviations from theoretical expectations in the relative intensities of the \amA\ main and satellite groups of hyperfine components. The observed anomalies in the NGC 1333 region do not support the theory involving hyperfine selective trapping \citep{Stutzki1985}. Although these anomalies affect the optical depth map, the effects on the temperature maps are negligible.

\item There is very good correlation of the \amA\ maps with the morphology and kinematics of the corresponding \nnh\ (J=1-0) maps. In regions away from the protostellar sources, the dust maps from JCMT and \textit{Herschel} also match well with the \amA\ integrated intensity map. They all trace the same material in these regions. 

\item The northern part of the region has many active outflows, which possibly cleared out a cavity between two filaments. 

\item In the southern part of the map, based on the continuous large velocity gradient pattern, the presence of Class 0/I YSOs along the region of the gradient and other kinematic signatures, we have argued that this is a region of compressed gas formed by the collision of large-scale turbulent cells.

\end{itemize} 

Overall, the \amm\ observations of the entire NGC 1333 region are very rich in information and offer many clues about star formation in molecular clouds. We have postulated theories to explain some of these observations. Further analysis of the data needs to be carried out to study the regions around each of the protostars, by investigating the relation between the various map features at all scales.

\begin{acknowledgments}

\end{acknowledgments}

\bibliographystyle{aasjournal}
\bibliography{refVLA}

\begin{thebibliography}{}
\expandafter\ifx\csname natexlab\endcsname\relax\def\natexlab#1{#1}\fi
\providecommand{\url}[1]{\href{#1}{#1}}

\bibitem[{{Alves} {et~al.}(2011){Alves}, {Acosta-Pulido}, {Girart}, {Franco},
  \& {L{\'o}pez}}]{Alves2011}
{Alves}, F.~O., {Acosta-Pulido}, J.~A., {Girart}, J.~M., {Franco}, G.~A.~P., \&
  {L{\'o}pez}, R. 2011, \aj, 142, 33

\bibitem[{{Andr{\'e}} {et~al.}(2010){Andr{\'e}}, {Men'shchikov}, {Bontemps},
  {K{\"o}nyves}, {Motte}, {Schneider}, {Didelon}, {Minier}, {Saraceno},
  {Ward-Thompson}, {di Francesco}, {White}, {Molinari}, {Testi}, {Abergel},
  {Griffin}, {Henning}, {Royer}, {Mer{\'{\i}}n}, {Vavrek}, {Attard},
  {Arzoumanian}, {Wilson}, {Ade}, {Aussel}, {Baluteau}, {Benedettini},
  {Bernard}, {Blommaert}, {Cambr{\'e}sy}, {Cox}, {di Giorgio}, {Hargrave},
  {Hennemann}, {Huang}, {Kirk}, {Krause}, {Launhardt}, {Leeks}, {Le Pennec},
  {Li}, {Martin}, {Maury}, {Olofsson}, {Omont}, {Peretto}, {Pezzuto}, {Prusti},
  {Roussel}, {Russeil}, {Sauvage}, {Sibthorpe}, {Sicilia-Aguilar}, {Spinoglio},
  {Waelkens}, {Woodcraft}, \& {Zavagno}}]{Andre2010}
{Andr{\'e}}, P., {Men'shchikov}, A., {Bontemps}, S., {et~al.} 2010, \aap, 518,
  L102

\bibitem[{{Arce}(2002)}]{ArceThesis}
{Arce}, H.~G. 2002, PhD thesis, Harvard University

\bibitem[{{Arce} {et~al.}(2011){Arce}, {Borkin}, {Goodman}, {Pineda}, \&
  {Beaumont}}]{Arce2011}
{Arce}, H.~G., {Borkin}, M.~A., {Goodman}, A.~A., {Pineda}, J.~E., \&
  {Beaumont}, C.~N. 2011, \apj, 742, 105

\bibitem[{{Balfour} {et~al.}(2015){Balfour}, {Whitworth}, {Hubber}, \&
  {Jaffa}}]{Balfour2015}
{Balfour}, S.~K., {Whitworth}, A.~P., {Hubber}, D.~A., \& {Jaffa}, S.~E. 2015,
  \mnras, 453, 2471

\bibitem[{{Ballesteros-Paredes} {et~al.}(1999){Ballesteros-Paredes},
  {V{\'a}zquez-Semadeni}, \& {Scalo}}]{BP1999}
{Ballesteros-Paredes}, J., {V{\'a}zquez-Semadeni}, E., \& {Scalo}, J. 1999,
  \apj, 515, 286

\bibitem[{{Bally} {et~al.}(1996){Bally}, {Devine}, \& {Reipurth}}]{Bally1996}
{Bally}, J., {Devine}, D., \& {Reipurth}, B. 1996, \apj, 473, L49

\bibitem[{{Banerjee} {et~al.}(2006){Banerjee}, {Pudritz}, \&
  {Anderson}}]{Banerjee2006}
{Banerjee}, R., {Pudritz}, R.~E., \& {Anderson}, D.~W. 2006, \mnras, 373, 1091

\bibitem[{{Belloche} {et~al.}(2006){Belloche}, {Hennebelle}, \&
  {Andr{\'e}}}]{Belloche2006}
{Belloche}, A., {Hennebelle}, P., \& {Andr{\'e}}, P. 2006, \aap, 453, 145

\bibitem[{{Bergin} \& {Langer}(1997)}]{BL1997}
{Bergin}, E.~A., \& {Langer}, W.~D. 1997, \apj, 486, 316

\bibitem[{{Beuther} {et~al.}(2015){Beuther}, {Ragan}, {Johnston}, {Henning},
  {Hacar}, \& {Kainulainen}}]{Beu2015}
{Beuther}, H., {Ragan}, S.~E., {Johnston}, K., {et~al.} 2015, \aap, 584, A67

\bibitem[{{Boldyrev} {et~al.}(2002){Boldyrev}, {Nordlund}, \&
  {Padoan}}]{Boldyrev2002}
{Boldyrev}, S., {Nordlund}, {\r{A}}., \& {Padoan}, P. 2002, \apj, 573, 678

\bibitem[{{Carroll} {et~al.}(2009){Carroll}, {Frank}, {Blackman}, {Cunningham},
  \& {Quillen}}]{Carroll2009}
{Carroll}, J.~J., {Frank}, A., {Blackman}, E.~G., {Cunningham}, A.~J., \&
  {Quillen}, A.~C. 2009, \apj, 695, 1376

\bibitem[{{Chen} \& {Ostriker}(2012)}]{ChenOst2012}
{Chen}, C.-Y., \& {Ostriker}, E.~C. 2012, \apj, 744, 124

\bibitem[{{Chen} \& {Ostriker}(2014)}]{ChenOst2014}
---. 2014, \apj, 785, 69

\bibitem[{{Chen} \& {Ostriker}(2015)}]{ChenOst2015}
---. 2015, \apj, 810, 126

\bibitem[{{Chen} {et~al.}(2016){Chen}, {Di Francesco}, {Johnstone}, {Sadavoy},
  {Hatchell}, {Mottram}, {Kirk}, {Buckle}, {Berry}, {Broekhoven- Fiene},
  {Currie}, {Fich}, {Jenness}, {Nutter}, {Pattle}, {Pineda}, {Quinn}, {Salji},
  {Tisi}, {Hogerheijde}, {Ward-Thompson}, {Bastien}, {Bresnahan}, {Butner},
  {Chrysostomou}, {Coude}, {Davis}, {Drabek-Maunder}, {Duarte-Cabral}, {Fiege},
  {Friberg}, {Friesen}, {Fuller}, {Graves}, {Greaves}, {Gregson}, {Holland},
  {Joncas}, {Kirk}, {Knee}, {Mairs}, {Marsh}, {Matthews}, {Moriarty-Schieven},
  {Mowat}, {Pezzuto}, {Rawlings}, {Richer}, {Robertson}, {Rosolowsky},
  {Rumble}, {Schneider-Bontemps}, {Thomas}, {Tothill}, {Viti}, {White},
  {Wouterloot}, {Yates}, \& {Zhu}}]{Chen2016}
{Chen}, M. C.-Y., {Di Francesco}, J., {Johnstone}, D., {et~al.} 2016, \apj,
  826, 95

\bibitem[{{COMPLETE team}(2011)}]{COMPLETE2011}
{COMPLETE team}. 2011, All Perseus in Sub-millimeter continuum (850 microns),
  vV2,  Harvard Dataverse, doi:10904/10089.
\newblock \url{https://hdl.handle.net/10904/10089}

\bibitem[{{Curtis} {et~al.}(2010){Curtis}, {Richer}, {Swift}, \&
  {Williams}}]{Curtis2010}
{Curtis}, E.~I., {Richer}, J.~S., {Swift}, J.~J., \& {Williams}, J.~P. 2010,
  \mnras, 408, 1516

\bibitem[{{Dewangan} \& {Ojha}(2017)}]{Dewangan2017}
{Dewangan}, L.~K., \& {Ojha}, D.~K. 2017, \apj, 849, 65

\bibitem[{{Dhabal} {et~al.}(2018){Dhabal}, {Mundy}, {Rizzo}, {Storm}, \&
  {Teuben}}]{Dhabal2018}
{Dhabal}, A., {Mundy}, L.~G., {Rizzo}, M.~J., {Storm}, S., \& {Teuben}, P.
  2018, \apj, 853, 169

\bibitem[{{Dionatos} \& {G{\"u}del}(2017)}]{Dionatus2017}
{Dionatos}, O., \& {G{\"u}del}, M. 2017, \aap, 597, A64

\bibitem[{{Dodds} {et~al.}(2015){Dodds}, {Greaves}, {Scholz}, {Hatchell},
  {Holland}, \& {JCMT Gould Belt Survey Team}}]{Dodds2015}
{Dodds}, P., {Greaves}, J.~S., {Scholz}, A., {et~al.} 2015, \mnras, 447, 722

\bibitem[{{Duarte-Cabral} {et~al.}(2011){Duarte-Cabral}, {Dobbs}, {Peretto}, \&
  {Fuller}}]{DC2011}
{Duarte-Cabral}, A., {Dobbs}, C.~L., {Peretto}, N., \& {Fuller}, G.~A. 2011,
  \aap, 528, A50

\bibitem[{{Dunham} {et~al.}(2015){Dunham}, {Allen}, {Evans},
  {Broekhoven-Fiene}, {Cieza}, {di Francesco}, {Gutermuth}, {Harvey},
  {Hatchell}, {Heiderman}, {Huard}, {Johnstone}, {Kirk}, {Matthews}, {Miller},
  {Peterson}, \& {Young}}]{Cat2015}
{Dunham}, M.~M., {Allen}, L.~E., {Evans}, II, N.~J., {et~al.} 2015, VizieR
  Online Data Catalog, 222

\bibitem[{{Elitzur} {et~al.}(1989){Elitzur}, {Hollenbach}, \&
  {McKee}}]{Elitzur1989}
{Elitzur}, M., {Hollenbach}, D.~J., \& {McKee}, C.~F. 1989, \apj, 346, 983

\bibitem[{{Flower}(2000)}]{Flower2000}
{Flower}, D.~R. 2000, \mnras, 313, L19

\bibitem[{{Forbrich} {et~al.}(2014){Forbrich}, {{\"O}berg}, {Lada}, {Lombardi},
  {Hacar}, {Alves}, \& {Rathborne}}]{Forbrich2014}
{Forbrich}, J., {{\"O}berg}, K., {Lada}, C.~J., {et~al.} 2014, \aap, 568, A27

\bibitem[{{Foster} {et~al.}(2009){Foster}, {Rosolowsky}, {Kauffmann}, {Pineda},
  {Borkin}, {Caselli}, {Myers}, \& {Goodman}}]{Foster2009}
{Foster}, J.~B., {Rosolowsky}, E.~W., {Kauffmann}, J., {et~al.} 2009, \apj,
  696, 298

\bibitem[{Friesen(2017)}]{GBT2017}
Friesen, R. 2017, NGC 1333 DR1 Data, vV1,  Harvard Dataverse,
  doi:10.7910/DVN/IP1ZZL.
\newblock \url{https://doi.org/10.7910/DVN/IP1ZZL}

\bibitem[{{Friesen} {et~al.}(2017){Friesen}, {Pineda}, {co-PIs}, {Rosolowsky},
  {Alves}, {Chac{\'o}n-Tanarro}, {How-Huan Chen}, {Chun-Yuan Chen}, {Di
  Francesco}, {Keown}, {Kirk}, {Punanova}, {Seo}, {Shirley}, {Ginsburg},
  {Hall}, {Offner}, {Singh}, {Arce}, {Caselli}, {Goodman}, {Martin}, {Matzner},
  {Myers}, {Redaelli}, \& {The GAS Collaboration}}]{Friesen2017}
{Friesen}, R.~K., {Pineda}, J.~E., {co-PIs}, {et~al.} 2017, \apj, 843, 63

\bibitem[{{Goldsmith} \& {Langer}(1999)}]{Goldsmith1999}
{Goldsmith}, P.~F., \& {Langer}, W.~D. 1999, \apj, 517, 209

\bibitem[{{Hacar} {et~al.}(2013){Hacar}, {Tafalla}, {Kauffmann}, \&
  {Kov{\'a}cs}}]{Hacar2013}
{Hacar}, A., {Tafalla}, M., {Kauffmann}, J., \& {Kov{\'a}cs}, A. 2013, \aap,
  554, A55

\bibitem[{{Hatchell} {et~al.}(2007){Hatchell}, {Fuller}, {Richer}, {Harries},
  \& {Ladd}}]{Hatchell2007}
{Hatchell}, J., {Fuller}, G.~A., {Richer}, J.~S., {Harries}, T.~J., \& {Ladd},
  E.~F. 2007, \aap, 468, 1009

\bibitem[{{Helfer} {et~al.}(2003){Helfer}, {Thornley}, {Regan}, {Wong},
  {Sheth}, {Vogel}, {Blitz}, \& {Bock}}]{Helfer2003}
{Helfer}, T.~T., {Thornley}, M.~D., {Regan}, M.~W., {et~al.} 2003, The
  Astrophysical Journal Supplement Series, 145, 259

\bibitem[{{Hennebelle} \& {Falgarone}(2012)}]{Henn2012}
{Hennebelle}, P., \& {Falgarone}, E. 2012, Astronomy and Astrophysics Review,
  20, 55

\bibitem[{{Hennebelle} {et~al.}(2003){Hennebelle}, {Whitworth}, {Gladwin}, \&
  {Andr{\'e}}}]{Henn2003}
{Hennebelle}, P., {Whitworth}, A.~P., {Gladwin}, P.~P., \& {Andr{\'e}}, P.
  2003, \mnras, 340, 870

\bibitem[{{Ho} \& {Townes}(1983)}]{Ho1983}
{Ho}, P.~T.~P., \& {Townes}, C.~H. 1983, Annual Review of Astronomy and
  Astrophysics, 21, 239

\bibitem[{{Hollenbach} {et~al.}(2013){Hollenbach}, {Elitzur}, \&
  {McKee}}]{Hollenbach2013}
{Hollenbach}, D., {Elitzur}, M., \& {McKee}, C.~F. 2013, \apj, 773, 70

\bibitem[{{Inoue} \& {Fukui}(2013)}]{InFuk2013}
{Inoue}, T., \& {Fukui}, Y. 2013, \apj, 774, L31

\bibitem[{{Johnstone} {et~al.}(2010){Johnstone}, {Rosolowsky}, {Tafalla}, \&
  {Kirk}}]{Johnstone2010}
{Johnstone}, D., {Rosolowsky}, E., {Tafalla}, M., \& {Kirk}, H. 2010, \apj,
  711, 655

\bibitem[{{J{\o}rgensen} {et~al.}(2006){J{\o}rgensen}, {Harvey}, {Evans},
  {Huard}, {Allen}, {Porras}, {Blake}, {Bourke}, {Chapman}, {Cieza}, {Koerner},
  {Lai}, {Mundy}, {Myers}, {Padgett}, {Rebull}, {Sargent}, {Spiesman},
  {Stapelfeldt}, {van Dishoeck}, {Wahhaj}, \& {Young}}]{Jorgensen2006}
{J{\o}rgensen}, J.~K., {Harvey}, P.~M., {Evans}, Neal~J., I., {et~al.} 2006,
  \apj, 645, 1246

\bibitem[{{Kirk} {et~al.}(2017){Kirk}, {Friesen}, {Pineda}, {Rosolowsky},
  {Offner}, {Matzner}, {Myers}, {Di Francesco}, {Caselli}, {Alves},
  {Chac{\'o}n-Tanarro}, {Chen}, {Chun-Yuan Chen}, {Keown}, {Punanova}, {Seo},
  {Shirley}, {Ginsburg}, {Hall}, {Singh}, {Arce}, {Goodman}, {Martin}, \&
  {Redaelli}}]{Kirk2017}
{Kirk}, H., {Friesen}, R.~K., {Pineda}, J.~E., {et~al.} 2017, \apj, 846, 144

\bibitem[{{Klessen} {et~al.}(2005){Klessen}, {Ballesteros-Paredes},
  {V{\'a}zquez-Semadeni}, \& {Dur{\'a}n-Rojas}}]{Klessen2005}
{Klessen}, R.~S., {Ballesteros-Paredes}, J., {V{\'a}zquez-Semadeni}, E., \&
  {Dur{\'a}n-Rojas}, C. 2005, \apj, 620, 786

\bibitem[{{Knee} \& {Sandell}(2000)}]{Knee2000}
{Knee}, L.~B.~G., \& {Sandell}, G. 2000, \aap, 361, 671

\bibitem[{{Koda} {et~al.}(2011){Koda}, {Sawada}, {Wright}, {Teuben}, {Corder},
  {Patience}, {Scoville}, {Donovan Meyer}, \& {Egusa}}]{Koda2011}
{Koda}, J., {Sawada}, T., {Wright}, M. C.~H., {et~al.} 2011, The Astrophysical
  Journal Supplement Series, 193, 19

\bibitem[{Kogan(2000)}]{Kogan2000}
Kogan, L. 2000, IEEE Transactions on Antennas and Propagation, 48, 1075

\bibitem[{{Krumholz}(2006)}]{Krumholz2006}
{Krumholz}, M.~R. 2006, \apj, 641, L45

\bibitem[{{Larson}(1981)}]{Larson1981}
{Larson}, R.~B. 1981, \mnras, 194, 809

\bibitem[{{Lefloch} {et~al.}(1998){Lefloch}, {Castets}, {Cernicharo}, {Langer},
  \& {Zylka}}]{Lefloch1998}
{Lefloch}, B., {Castets}, A., {Cernicharo}, J., {Langer}, W.~D., \& {Zylka}, R.
  1998, \aap, 334, 269

\bibitem[{{Li} \& {Nakamura}(2004)}]{Li2004}
{Li}, Z.-Y., \& {Nakamura}, F. 2004, \apjl, 609, L83

\bibitem[{{Lyo} {et~al.}(2014){Lyo}, {Kim}, {Byun}, \& {Lee}}]{Lyo2014}
{Lyo}, A.~R., {Kim}, J., {Byun}, D.-Y., \& {Lee}, H.-G. 2014, \aj, 148, 80

\bibitem[{{Mac Low} \& {Klessen}(2004)}]{MLKlessen2004}
{Mac Low}, M.-M., \& {Klessen}, R.~S. 2004, Reviews of Modern Physics, 76, 125

\bibitem[{{Mangum} \& {Shirley}(2015)}]{Magnum2015}
{Mangum}, J.~G., \& {Shirley}, Y.~L. 2015, \pasp, 127, 266

\bibitem[{{Mangum} {et~al.}(1992){Mangum}, {Wootten}, \& {Mundy}}]{Magnum1992}
{Mangum}, J.~G., {Wootten}, A., \& {Mundy}, L.~G. 1992, \apj, 388, 467

\bibitem[{{Mestel} \& {Spitzer}(1956)}]{MS1956}
{Mestel}, L., \& {Spitzer}, Jr., L. 1956, \mnras, 116, 503

\bibitem[{{Mizuno} {et~al.}(1995){Mizuno}, {Onishi}, {Yonekura}, {Nagahama},
  {Ogawa}, \& {Fukui}}]{Mizuno1995}
{Mizuno}, A., {Onishi}, T., {Yonekura}, Y., {et~al.} 1995, \apj, 445, L161

\bibitem[{{Mouschovias}(1979)}]{Mous1979}
{Mouschovias}, T.~C. 1979, \apj, 228, 475

\bibitem[{{M{\"u}ller} {et~al.}(2005){M{\"u}ller}, {Schl{\"o}der}, {Stutzki},
  \& {Winnewisser}}]{CDMS2005}
{M{\"u}ller}, H. S.~P., {Schl{\"o}der}, F., {Stutzki}, J., \& {Winnewisser}, G.
  2005, Journal of Molecular Structure, 742, 215

\bibitem[{{Nakamura} \& {Li}(2005)}]{Nakamura2005}
{Nakamura}, F., \& {Li}, Z.-Y. 2005, \apj, 631, 411

\bibitem[{{Nakamura} {et~al.}(2014){Nakamura}, {Sugitani}, {Tanaka},
  {Nishitani}, {Dobashi}, {Shimoikura}, {Shimajiri}, {Kawabe}, {Yonekura},
  {Mizuno}, {Kimura}, {Tokuda}, {Kozu}, {Okada}, {Hasegawa}, {Ogawa}, {Kameno},
  {Shinnaga}, {Momose}, {Nakajima}, {Onishi}, {Maezawa}, {Hirota}, {Takano},
  {Iono}, {Kuno}, \& {Yamamoto}}]{Nakamura2014}
{Nakamura}, F., {Sugitani}, K., {Tanaka}, T., {et~al.} 2014, \apj, 791, L23

\bibitem[{{Offner} \& {Arce}(2015)}]{Offner2015}
{Offner}, S. S.~R., \& {Arce}, H.~G. 2015, \apj, 811, 146

\bibitem[{{Offner} {et~al.}(2009){Offner}, {Klein}, {McKee}, \&
  {Krumholz}}]{Offner2009}
{Offner}, S. S.~R., {Klein}, R.~I., {McKee}, C.~F., \& {Krumholz}, M.~R. 2009,
  \apj, 703, 131

\bibitem[{{Padoan} {et~al.}(1999){Padoan}, {Bally}, {Billawala}, {Juvela}, \&
  {Nordlund}}]{Padoan1999}
{Padoan}, P., {Bally}, J., {Billawala}, Y., {Juvela}, M., \& {Nordlund},
  {\r{A}}. 1999, \apj, 525, 318

\bibitem[{{Padoan} \& {Nordlund}(2002)}]{Padoan2002}
{Padoan}, P., \& {Nordlund}, {\r{A}}. 2002, \apj, 576, 870

\bibitem[{{Pattle} {et~al.}(2015){Pattle}, {Ward-Thompson}, {Kirk}, {White},
  {Drabek-Maunder}, {Buckle}, {Beaulieu}, {Berry}, {Broekhoven-Fiene},
  {Currie}, {Fich}, {Hatchell}, {Kirk}, {Jenness}, {Johnstone}, {Mottram},
  {Nutter}, {Pineda}, {Quinn}, {Salji}, {Tisi}, {Walker-Smith}, {di Francesco},
  {Hogerheijde}, {Andr{\'e}}, {Bastien}, {Bresnahan}, {Butner}, {Chen},
  {Chrysostomou}, {Coude}, {Davis}, {Duarte-Cabral}, {Fiege}, {Friberg},
  {Friesen}, {Fuller}, {Graves}, {Greaves}, {Gregson}, {Griffin}, {Holland},
  {Joncas}, {Knee}, {K{\"o}nyves}, {Mairs}, {Marsh}, {Matthews},
  {Moriarty-Schieven}, {Rawlings}, {Richer}, {Robertson}, {Rosolowsky},
  {Rumble}, {Sadavoy}, {Spinoglio}, {Thomas}, {Tothill}, {Viti}, {Wouterloot},
  {Yates}, \& {Zhu}}]{Pattle2015}
{Pattle}, K., {Ward-Thompson}, D., {Kirk}, J.~M., {et~al.} 2015, \mnras, 450,
  1094

\bibitem[{{Pezzuto} {et~al.}(2017){Pezzuto}, {Fiorellino}, {Benedettini},
  {Schisano}, {Elia}, {Andr{\'e}}, {K{\"o}nyves}, {Ladjelate}, {Di Francesco},
  {Piccotti}, \& {Herschel Gould Belt Survey Consortium}}]{Pezzuto2017}
{Pezzuto}, S., {Fiorellino}, E., {Benedettini}, M., {et~al.} 2017, Memorie
  della Societa Astronomica Italiana, 88, 806

\bibitem[{{Pineda} {et~al.}(2010){Pineda}, {Goodman}, {Arce}, {Caselli},
  {Foster}, {Myers}, \& {Rosolowsky}}]{Pineda2010}
{Pineda}, J.~E., {Goodman}, A.~A., {Arce}, H.~G., {et~al.} 2010, \apjl, 712,
  L116

\bibitem[{{Plunkett} {et~al.}(2013){Plunkett}, {Arce}, {Corder}, {Mardones},
  {Sargent}, \& {Schnee}}]{Plunkett2013}
{Plunkett}, A.~L., {Arce}, H.~G., {Corder}, S.~A., {et~al.} 2013, \apj, 774, 22

\bibitem[{{Pudritz} \& {Kevlahan}(2013)}]{PudKev2013}
{Pudritz}, R.~E., \& {Kevlahan}, N.~K.~R. 2013, Philosophical Transactions of
  the Royal Society of London Series A, 371, 20120248

\bibitem[{{Quillen} {et~al.}(2005){Quillen}, {Thorndike}, {Cunningham},
  {Frank}, {Gutermuth}, {Blackman}, {Pipher}, \& {Ridge}}]{Quillen2005}
{Quillen}, A.~C., {Thorndike}, S.~L., {Cunningham}, A., {et~al.} 2005, \apj,
  632, 941

\bibitem[{{Rosolowsky} {et~al.}(2008){Rosolowsky}, {Pineda}, {Foster},
  {Borkin}, {Kauffmann}, {Caselli}, {Myers}, \& {Goodman}}]{Rosolowsky2008}
{Rosolowsky}, E.~W., {Pineda}, J.~E., {Foster}, J.~B., {et~al.} 2008, The
  Astrophysical Journal Supplement Series, 175, 509

\bibitem[{{Sadavoy} \& {Stahler}(2017)}]{SS2017}
{Sadavoy}, S.~I., \& {Stahler}, S.~W. 2017, \mnras, 469, 3881

\bibitem[{{Sadavoy} {et~al.}(2014){Sadavoy}, {Di Francesco}, {Andr{\'e}},
  {Pezzuto}, {Bernard}, {Maury}, {Men'shchikov}, {Motte}, {Nguyen-Lu'o'ng},
  {Schneider}, {Arzoumanian}, {Benedettini}, {Bontemps}, {Elia}, {Hennemann},
  {Hill}, {K{\"o}nyves}, {Louvet}, {Peretto}, {Roy}, \& {White}}]{Sadavoy2014}
{Sadavoy}, S.~I., {Di Francesco}, J., {Andr{\'e}}, P., {et~al.} 2014, \apjl,
  787, L18

\bibitem[{{Sandell} \& {Knee}(2001)}]{Sandell2001}
{Sandell}, G., \& {Knee}, L. 2001, in Science with the Atacama Large Millimeter
  Array, Vol. 235, 154

\bibitem[{{Schnee} {et~al.}(2006){Schnee}, {Bethell}, \&
  {Goodman}}]{Schnee2006}
{Schnee}, S., {Bethell}, T., \& {Goodman}, A. 2006, \apj, 640, L47

\bibitem[{{Seo} {et~al.}(2015){Seo}, {Shirley}, {Goldsmith}, {Ward-Thompson},
  {Kirk}, {Schmalzl}, {Lee}, {Friesen}, {Langston}, {Masters}, \&
  {Garwood}}]{Seo2015}
{Seo}, Y.~M., {Shirley}, Y.~L., {Goldsmith}, P., {et~al.} 2015, \apj, 805, 185

\bibitem[{{Shirley}(2015)}]{Shirley2015}
{Shirley}, Y.~L. 2015, Publications of the Astronomical Society of the Pacific,
  127, 299

\bibitem[{{Shu}(1992)}]{Shu1992}
{Shu}, F.~H. 1992, {Physics of Astrophysics, Vol. II} (University Science
  Books)

\bibitem[{{Smith} {et~al.}(2014){Smith}, {Glover}, \& {Klessen}}]{Smith2014}
{Smith}, R.~J., {Glover}, S.~C.~O., \& {Klessen}, R.~S. 2014, \mnras, 445, 2900

\bibitem[{{Smith} {et~al.}(2016){Smith}, {Glover}, {Klessen}, \&
  {Fuller}}]{Smith2016}
{Smith}, R.~J., {Glover}, S.~C.~O., {Klessen}, R.~S., \& {Fuller}, G.~A. 2016,
  \mnras, 455, 3640

\bibitem[{{Spitzer}(1968)}]{Spitzer1968}
{Spitzer}, L. 1968, {Diffuse matter in space} (New York: Interscience
  Publication, 1968)

\bibitem[{{Storm} {et~al.}(2014){Storm}, {Mundy}, {Fern{\'a}ndez-L{\'o}pez},
  {Lee}, {Looney}, {Teuben}, {Rosolowsky}, {Arce}, {Ostriker}, {Segura-Cox},
  {Pound}, {Salter}, {Volgenau}, {Shirley}, {Chen}, {Gong}, {Plunkett},
  {Tobin}, {Kwon}, {Isella}, {Kauffmann}, {Tassis}, {Crutcher}, {Gammie}, \&
  {Testi}}]{Storm2014}
{Storm}, S., {Mundy}, L.~G., {Fern{\'a}ndez-L{\'o}pez}, M., {et~al.} 2014,
  \apj, 794, 165

\bibitem[{{Stutzki} {et~al.}(1984){Stutzki}, {Jackson}, {Olberg}, {Barrett}, \&
  {Winnewisser}}]{Stutzki1984}
{Stutzki}, J., {Jackson}, J.~M., {Olberg}, M., {Barrett}, A.~H., \&
  {Winnewisser}, G. 1984, \aap, 139, 258

\bibitem[{{Stutzki} \& {Winnewisser}(1985)}]{Stutzki1985}
{Stutzki}, J., \& {Winnewisser}, G. 1985, \aap, 144, 13

\bibitem[{{Swift} {et~al.}(2005){Swift}, {Welch}, \& {Di
  Francesco}}]{Swift2005}
{Swift}, J.~J., {Welch}, W.~J., \& {Di Francesco}, J. 2005, \apj, 620, 823

\bibitem[{{Tafalla} \& {Hacar}(2015)}]{Taf2015}
{Tafalla}, M., \& {Hacar}, A. 2015, \aap, 574, A104

\bibitem[{{Tafalla} \& {Myers}(1997)}]{Taf1997}
{Tafalla}, M., \& {Myers}, P.~C. 1997, \apj, 491, 653

\bibitem[{{Takahira} {et~al.}(2014){Takahira}, {Tasker}, \&
  {Habe}}]{Takahira2014}
{Takahira}, K., {Tasker}, E.~J., \& {Habe}, A. 2014, \apj, 792, 63

\bibitem[{{Teuben} {et~al.}(2018){Teuben}, {Burkutean}, \&
  {Stanke}}]{Teuben2018}
{Teuben}, P., {Burkutean}, S., \& {Stanke}, T. 2018, in Submillimetre
  Single-dish Data Reduction and Array Combination Techniques, 7

\bibitem[{{Tobin} {et~al.}(2015){Tobin}, {Dunham}, {Looney}, {Li}, {Chandler},
  {Segura-Cox}, {Sadavoy}, {Melis}, {Harris}, {Perez}, {Kratter},
  {J{\o}rgensen}, {Plunkett}, \& {Hull}}]{Tobin2015}
{Tobin}, J.~J., {Dunham}, M.~M., {Looney}, L.~W., {et~al.} 2015, \apj, 798, 61

\bibitem[{{V{\'a}zquez-Semadeni} {et~al.}(2003){V{\'a}zquez-Semadeni},
  {Ballesteros-Paredes}, \& {Klessen}}]{VS2003}
{V{\'a}zquez-Semadeni}, E., {Ballesteros-Paredes}, J., \& {Klessen}, R. 2003,
  in Galactic Star Formation Across the Stellar Mass Spectrum, Vol. 287, 81--86

\bibitem[{{V{\'a}zquez-Semadeni} {et~al.}(2011){V{\'a}zquez-Semadeni},
  {Banerjee}, {G{\'o}mez}, {Hennebelle}, {Duffin}, \& {Klessen}}]{VS2011}
{V{\'a}zquez-Semadeni}, E., {Banerjee}, R., {G{\'o}mez}, G.~C., {et~al.} 2011,
  \mnras, 414, 2511

\bibitem[{{Vogel} {et~al.}(1984){Vogel}, {Wright}, {Plambeck}, \&
  {Welch}}]{Vogel1984}
{Vogel}, S.~N., {Wright}, M.~C.~H., {Plambeck}, R.~L., \& {Welch}, W.~J. 1984,
  \apj, 283, 655

\bibitem[{{Volgenau} {et~al.}(2006){Volgenau}, {Mundy}, {Looney}, \&
  {Welch}}]{Volgenau2006}
{Volgenau}, N.~H., {Mundy}, L.~G., {Looney}, L.~W., \& {Welch}, W.~J. 2006,
  \apj, 651, 301

\bibitem[{{Walawender} {et~al.}(2008){Walawender}, {Bally}, {Francesco},
  {J{\o}rgensen}, \& {Getman}}]{Walawender2008}
{Walawender}, J., {Bally}, J., {Francesco}, J.~D., {J{\o}rgensen}, J., \&
  {Getman}, K.~. 2008, {NGC 1333: A Nearby Burst of Star Formation} (The
  Northern Sky ASP Monograph Publications), 346

\bibitem[{{Walawender} {et~al.}(2005){Walawender}, {Bally}, \&
  {Reipurth}}]{Walawender2005}
{Walawender}, J., {Bally}, J., \& {Reipurth}, B. 2005, \aj, 129, 2308

\bibitem[{{Wei{\ss}} {et~al.}(2001){Wei{\ss}}, {Neininger}, {H{\"u}ttemeister},
  \& {Klein}}]{Weiss2001}
{Wei{\ss}}, A., {Neininger}, N., {H{\"u}ttemeister}, S., \& {Klein}, U. 2001,
  \aap, 365, 571

\bibitem[{{Young} {et~al.}(2015){Young}, {Young}, {Lai}, {Dunham}, \&
  {Evans}}]{Young2015}
{Young}, K.~E., {Young}, C.~H., {Lai}, S.-P., {Dunham}, M.~M., \& {Evans},
  Neal~J., I. 2015, \aj, 150, 40

\bibitem[{{Zucker} {et~al.}(2018){Zucker}, {Schlafly}, {Speagle}, {Green},
  {Portillo}, {Finkbeiner}, \& {Goodman}}]{Zucker2018}
{Zucker}, C., {Schlafly}, E.~F., {Speagle}, J.~S., {et~al.} 2018, \apj, 869, 83

\end{thebibliography}

\appendix
\section{Weighting of Single Dish and Interferometric Visibilities}
\label{sec:AppA}

The value of the weighting parameter used prior to visibility combination in Section \ref{sec:Comb} involves matching the weight densities of the single dish visibilities and the interferometric visibilities. Since the single dish weight density falls off rapidly with uv distance, the matching is not very exact. In this section, we investigate the sensitivity of the maps and the total flux with the relative weighting. 

\begin{figure*}
\centering
\includegraphics[width=0.9\textwidth]{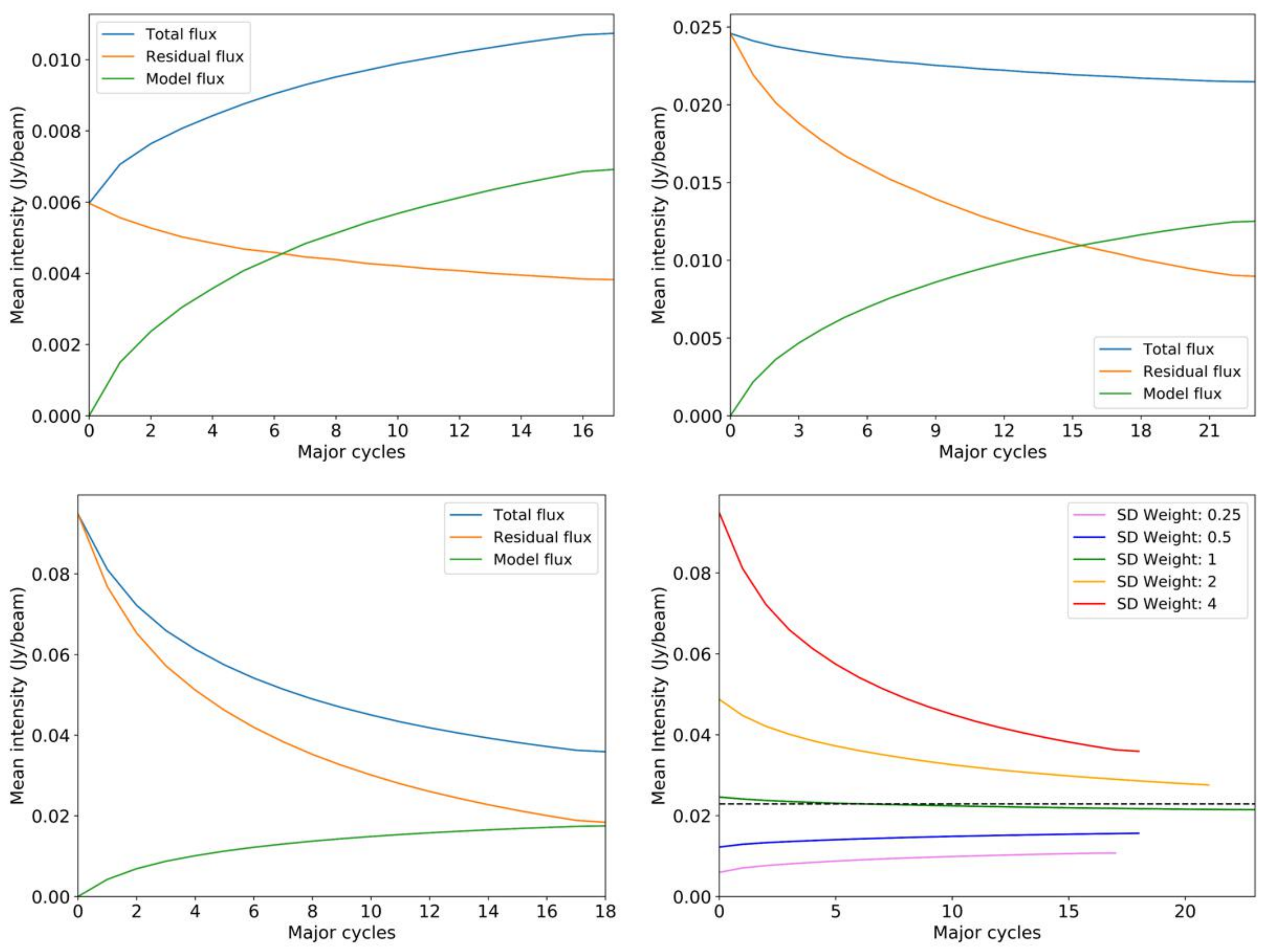}
\caption{Evolution of model flux, residual flux and total flux with major cycles of cleaning iterations for a single channel (8.11-8.25~km/s) sub-region containing signal for VLA and GBT combined dirty maps. \textit{Top-left, top-right and bottom-left}: Flux evolution for different weights of the GBT visibilities (0.25, 1 and 4 times the nominal weights, respectively) used while combining with the VLA visibilities. \textit{Bottom-right}: Comparison of the total flux evolution with cleaning for different weights of GBT visibilities. The dotted line indicates the total flux in the corresponding GBT map channel (scaled based on the beam size ratio). The total number of iterations is different in each case because the threshold limit of 3$\sigma$ (where $\sigma$ is the RMS noise of the combined dirty map) is reached after different number of major cycles.}
\label{fig:WtComp1}
\end{figure*}

\begin{figure*}
\centering
\includegraphics[trim={2.5cm 2.55cm 2.3cm 3.6cm},clip,width=\textwidth]{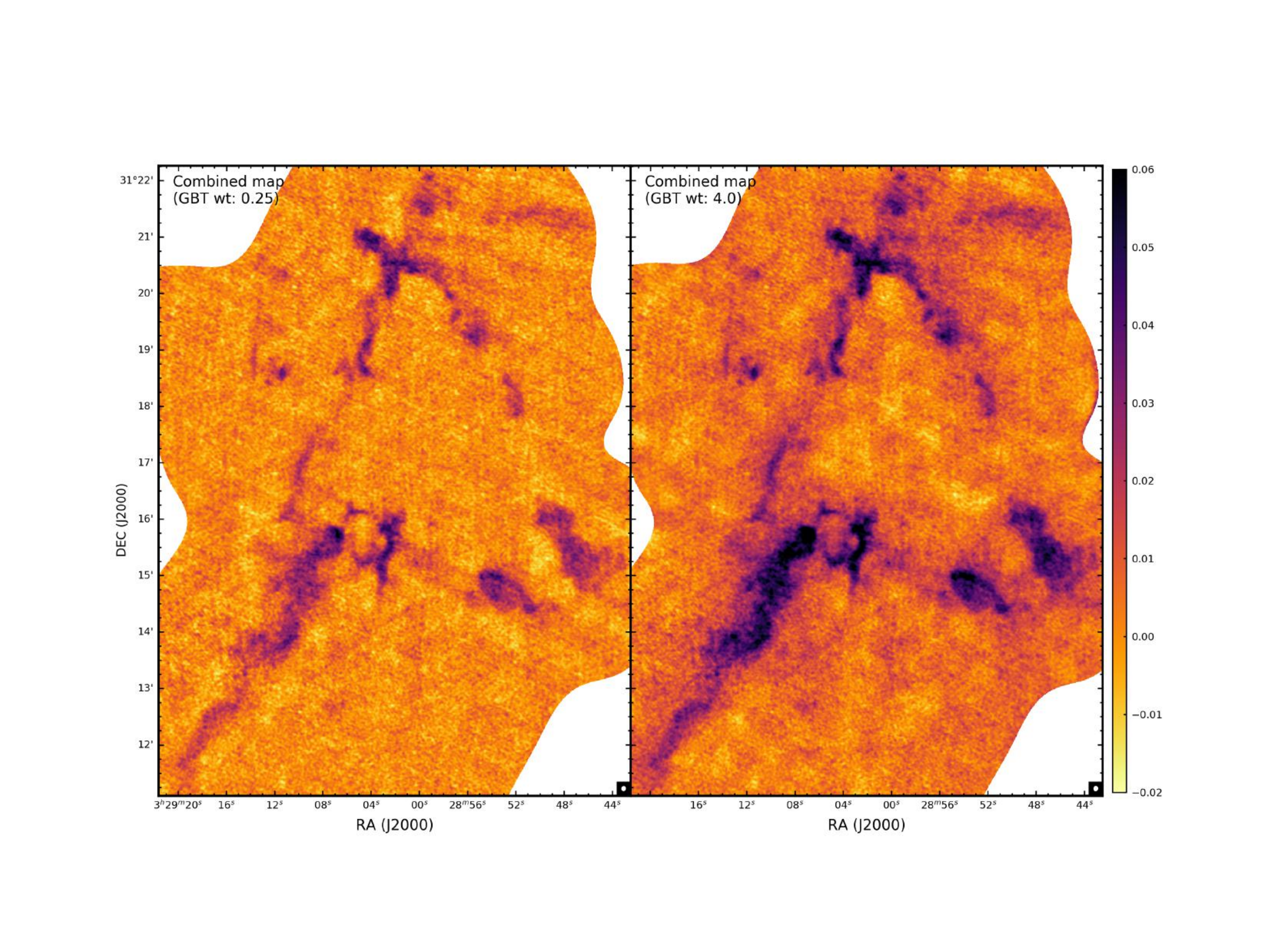}
\caption{Comparisons of combined VLA and GBT maps for a single channel (8.11-8.25~km/s) for different weights of the GBT visibilities: 0.25 (left) and 4.0 (right) times the nominal weight. The colorbar is in Jy/beam. The left image shows more of the interferometric artifacts, while in the right image the emission is more extended and the total flux is overestimated by about a factor of 2.}
\label{fig:WtComp2}
\end{figure*}

Assuming the GBT weighting parameter that is used for the data analysis to be 1, we use GBT weights in the range of 0.25 to 4 on a single channel containing signal to obtain the corresponding dirty maps and study the progression of the cleaning process in each of them. We used the same `cycleniter' parameter as before (1000), and set a threshold of 3 times the RMS noise in the dirty maps. As shown in Figure \ref{fig:WtComp1}, in the nominal case the total flux remains close to the GBT flux (scaled for the beam area) throughout the cleaning cycles. For higher (lower) than nominal GBT weights, the total flux in the dirty map is very high (low), but with progressive iterations the final cleaned flux reaches closer to that of the nominal weight case. Eventually, they end up within a factor of 2 of the total flux in the nominal case for a weight factor of 4. Most of the contribution to this difference comes from the extent of the emission, which increases with increasing GBT weights. The differences in regions of peak emission are lesser ($<$ 10\% for a weight factor of 4). The model flux and residual flux also end up at different ratios with respect to the total flux in the different cases. The beam areas only vary by about 2\% for a factor of 4 in the weights. We compare the outcomes of the final cleaned channel maps for GBT weights 0.25 and 4 in Figure \ref{fig:WtComp2}.

\end{document}